\documentclass{article}

\usepackage{arxiv}

\usepackage[utf8]{inputenc} % allow utf-8 input
\usepackage[T1]{fontenc}    % use 8-bit T1 fonts
\usepackage{hyperref}       % hyperlinks
\usepackage{url}            % simple URL typesetting
\usepackage{booktabs}       % professional-quality tables
\usepackage{nicefrac}       % compact symbols for 1/2, etc.
\usepackage{microtype}      % microtypography
\usepackage{lipsum}
\usepackage{fancyhdr}       % header
\usepackage{xcolor}

\usepackage{color}
\usepackage{graphicx}
\usepackage{float}
\usepackage{subfloat}

\usepackage{subfigure}
\usepackage{epsfig}
\usepackage{psfrag}
\usepackage{tabulary}

\usepackage[english]{babel}
\usepackage{ifthen,shortvrb}
\usepackage[fleqn]{amsmath}
\usepackage{amssymb}

\usepackage{graphicx}
\usepackage{dcolumn}

\usepackage{bm}
\usepackage{amsfonts,amsmath,amssymb}

\usepackage{amsthm}

\theoremstyle{definition}

\theoremstyle{remark}

\usepackage{natbib}
\usepackage{amssymb}
\usepackage{geometry}
\usepackage{epstopdf}

\usepackage{lipsum}

\usepackage{color} 
\usepackage{listings} 
\usepackage{setspace} 
 
\usepackage{framed}

\usepackage{caption}

\usepackage{algorithm}
\usepackage{algpseudocode}

\title{Computational modeling of viscoelastic backsheet materials for photovoltaics
}

\author{
  Ajinkya R. Dusane \\
  Research Unit Multi-scale Analysis of Materials (MUSAM),
  IMT School for Advanced Studies Lucca, Italy. \\
  \texttt{ajinkya.dusane@imtlucca.it} \\
\And
   Pietro Lenarda \\
  Research Unit Multi-scale Analysis of Materials (MUSAM),
  IMT School for Advanced Studies Lucca, Italy. \\
  \texttt{pietro.lenarda@imtlucca.it} \\
  \And
  Marco Paggi \\
  Research Unit Multi-scale Analysis of Materials (MUSAM),
  IMT School for Advanced Studies Lucca, Italy. \\
  \texttt{marco.paggi@imtlucca.it} \\
}

\begin{document}
\maketitle
\begin{abstract}
Backsheet is the outermost layer of the photovoltaic (PV) laminate which consists of polymers such as Polyethylene terephthalate (PET), Polyvinyl fluoride (PVF).
The viscoelastic response of these materials significantly affects the durability of the PV module.
In this study, the viscoelastic response of commercially available backsheet materials is experimentally characterized and computationally modeled.
An extensive viscoelastic experimental study on backsheet materials is carried out, considering the temperature-dependent properties for characterizing the mechanical properties. 
Based on an experimental campaign, 
small-strain viscoelastic models based on the Prony-series (PS) and Fractional Calculus (FC) Models are proposed here.
The form of the constitutive equations for both models is summarized, and the finite element implementation is described in detail.
%A novel finite element formulation and numerical implementation are proposed for the simulation of transient thermal analysis of backsheets.
Following identifications of relevant material parameters, we validate the model with the experimental data that shows good predictability. 
A comparative study of model responses under different loading conditions is also reported to assess the advantages and disadvantages of both models.
Such an extensive experimental study and constitutive modeling will help design and simulate a more comprehensive modeling of PV modules, as illustrated by the benchmark problems.
\end{abstract}

% keywords can be removed

\section{Introduction}
\begin{figure}[h!]
	\centering
	\includegraphics[width=0.50\linewidth]{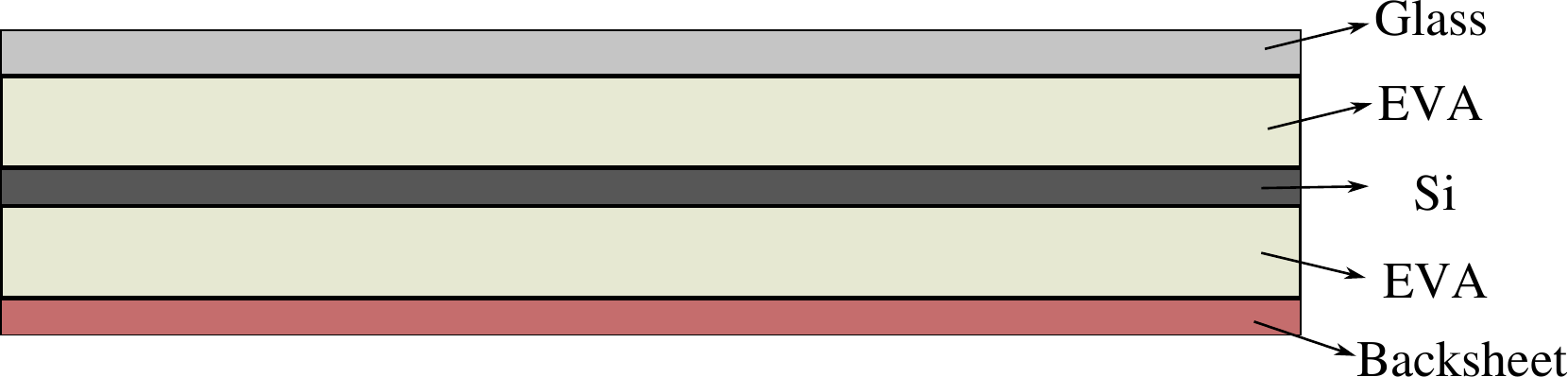}
	\caption{Schematic representing different components of a PV module. \hspace{40pt} }
	\label{solcomp}
\end{figure}

Photovoltaic (PV) modules are laminates composed of thin layers, as schematically shown in Fig. \ref{solcomp}. 
The durability of PV modules has always been a subject of concern 
\citep{en14144278, paggi2016global}, primarily because of their multilayered structure. 
Various material degradation mechanisms have been discussed in the literature to understand the origin of the PV module's overall underperformance \citep{liu2019quantitative}.
The mechanical and electrical response of PV modules may degrade over time due to various environmental factors, such as humidity, solar radiation, temperature, and external loading, causing mechanical damage in Silicon solar cells in the polymeric materials \citep{vazquez2008photovoltaic,borri2018fatigue, LIU2022108125}. 
Developing a comprehensive multiphysics model that can predict the overall mechanical behavior of PV modules under different conditions is still an active area of research \citep{omazic2019relation}. 
Till now, the scientific community has given more attention to EVA (Poly-ethylene Vinyl Acetate), which encapsulates the silicon solar cells \citep{CZANDERNA1996101}. 
The EVA mechanical properties, its chemical and physical degradation, and its response to temperature are widely known \citep{de2018causes,gagliardi2017reaction} and, in most cases, they have been successfully modeled \citep{hirschl2013determining}.
It can now be predicted with high accuracy. 
On the other hand, much less attention has been given to the experimental characterization and modeling of the backsheet materials, which are used for the outermost layer of the PV module. 
Several commercial types are available with very different physico-mechanical characteristics. 
The primary purpose of the backsheet is to protect the inner components of the module, specifically the Silicon solar cells and electric components, from external loading and also to act as an electric insulator. Therefore, any damage may cause a severe safety hazard.
Usually, backsheets are made up of polymers such as Polyethylene terephthalate (PET), Polyvinyl fluoride (PVF), Polyvinylidene fluoride (PVDF), and Polyamide (PA), with thicknesses ranging from 30 to 270 $\mu$m. 
\\
Analysis of the backsheet mainly emphasized material identification of polymers to be incorporated into the PV module and the determination of material degradation effects/failure analysis, which are based on chemical and physical material analysis of their constituent polymers. These materials often show a complex rheological behavior, changing their properties with time. 
In most cases, responses of backsheets are modeled within the elastic regime only \citep{ottersbock2022accelerate}. 
At the same time, there is a lack of a complete characterization of their viscoelastic response. 
The elastic modulus of the backsheet material may vary depending on the temperature and corresponding relaxation time. 
Similar to EVA and other polymers, backsheet materials also show power-law-type stress relaxation, and their behavior can be studied through viscoelasticity. 
Backsheet properties significantly influence the stress and deformation state of the PV module \citep{dietrich2010mechanical}. 
In order to identify the viscoelastic response of the material, various experimental tests, such as relaxation tests (at a constant applied strain level, stress is recorded) and creep tests (at a constant applied stress level, strain is recorded). 
Apart from these, Dynamic-Mechanical Analysis (DMA tests) can be used to determine the characteristics of materials efficiently. 
The latter applies sinusoidal-varying stress and measures the deformation of the material, allowing the determination of the storage modulus as well as the loss modulus. Based on frequency response, viscoelastic material parameters are identified \citep{bosco2020viscoelastic}.
The properties of viscoelastic materials strongly depend on temperature as well. The temperature dependency can be taken into account by means of the Time-Temperature-Superposition-Principle (TTSP) \citep{kraus2017parameter}, which provides all relaxation/creep functions at a given temperature from the material response at a reference temperature, $T_{ref}$. 
In practice, back sheets are often exposed to high temperatures with a load that induces constant stress or strain over a period of time. 
Therefore, relaxation tests seem to be the most appropriate test for identifying the parameters. 
Various mathematical models have been proposed in the literature to model the viscoelastic response of materials \citep{taylor1970thermomechanical,knauss2015review}. 
The primary aim of modeling is to extract model parameters and relate them to different deformation mechanisms. 
The linear viscoelastic model is an integro-differential equation in which the elastic stress tensor is a convolution product between the strain with a time-dependent elastic relaxation modulus. 
Models based on a combination of springs and dashpots are often used to get exponential-type relations and provide an approximate power-law trend of the elastic parameters' dependence over time. A more versatile Maxwell model proposed by Maxwell and Wiechart \citep{marques2012computational} consists of a linear spring and a dashpot in parallel, giving rise to a Generalized Maxwell Model (GMM), also called Prony Series (PS) model, which is used to fit experimental data \citep{kaliske1997formulation,reese1998theory,simo1987fully}. 
A high number of prony series is often required to improve the fitting accuracy, giving rise to several drawbacks, including the fact that many unknown parameters must be identified \citep{pander2011thermo}. 
The PS models are already in use to describe the linear viscoelastic behavior of solids such as polymers \citep{xu2018mathematical,eitner2011thermomechanics,sanchez2022numerical}, asphalt concrete\citep{xu2009modelling}, biological materials \cite{budday2017viscoelastic}, etc. 
The numerical implementation of the PS models is also computationally convenient due to its exponential format for closed-form solutions in time integration \cite{chen2000determining,reese1998theory}. 
Alternatively, models based on fractional calculus (FC) are becoming a promising alternative to modeling the power-law dependency of the relaxation behavior of polymers 
\citep{paggi2015accurate}. As power-law relations rise naturally by assuming a material constitutive law of fractional type, that is, 
involving noninteger order derivatives of stress and/or strain \citep{alotta2018finite, bonfanti2020fractional}.
Implementation of FC models is pretty straightforward, as convolution integral can be represented in the form of Caputo fractional derivative \citep{samko1993fractional,alotta2018viscoelastic}.
Recent work by \cite{lenarda2022computational} shows that fractional calculus offers the easiest way to estimate the model parameters compared to other models. 
\\
Based on these premises, the contribution of the current study has the following specific aims. At first, a comprehensive experimental campaign on different types of commercial backsheets to evaluate their material properties, considering uniaxial tensile tests and temperature-dependent relaxation tests, was performed. 
We have proposed small-strain linear viscoelastic models based on prony series and fractional calculus to model the viscoelastic response. So, a valid comparative study can be made. 
The form of the constitutive equations for both models is summarized in brief, and the finite element implementation  is described in detail. 
Since experimental results are obtained using uniaxial tests, we reformulate the two-dimensional viscoelastic constitutive model to one-dimensional forms.
Afterward, all parameters of the models are identified using an optimization procedure with additional focus given to addressing the issue of reducing model parameters. 
For that reason, a MATLAB program has been written for parameter identification. Finally, some benchmark problems are discussed to show the capability of the developed models through Finite Element Analysis (FEA). 
The manuscript is organized as follows. A brief note describing experimental tests with results carried out in the MUSAM-Lab at the IMT School for Advanced Studies Lucca is presented in Sec. \ref{experimental tests}. 
Sec. \ref{Matmodels} provides an overview of the linear viscoelastic material models herein considered. The proposed prony series (PS) and the fractional calculus (FC) models are described in detail, and the complexities arising from their parameter identification are discussed. 
Moreover, their finite element implementation is described. Sec. \ref{Paraidentify} discusses the proposed optimization procedure for parameter identification and presents the results for the considered backsheets, with a critical perspective regarding the two different viscoelastic models. 
Sec. \ref{numericalana} presents the numerical analysis and model validations with previously identified parameters. Finally, concluding remarks close the manuscript in Sec. \ref{conclusion}, with an outlook on future developments.

\section{Experimental tests}
\label{experimental tests}
\subsection{Material description and properties}
\begin{figure}[h!]
    \centering
    \includegraphics[width=0.95\linewidth]{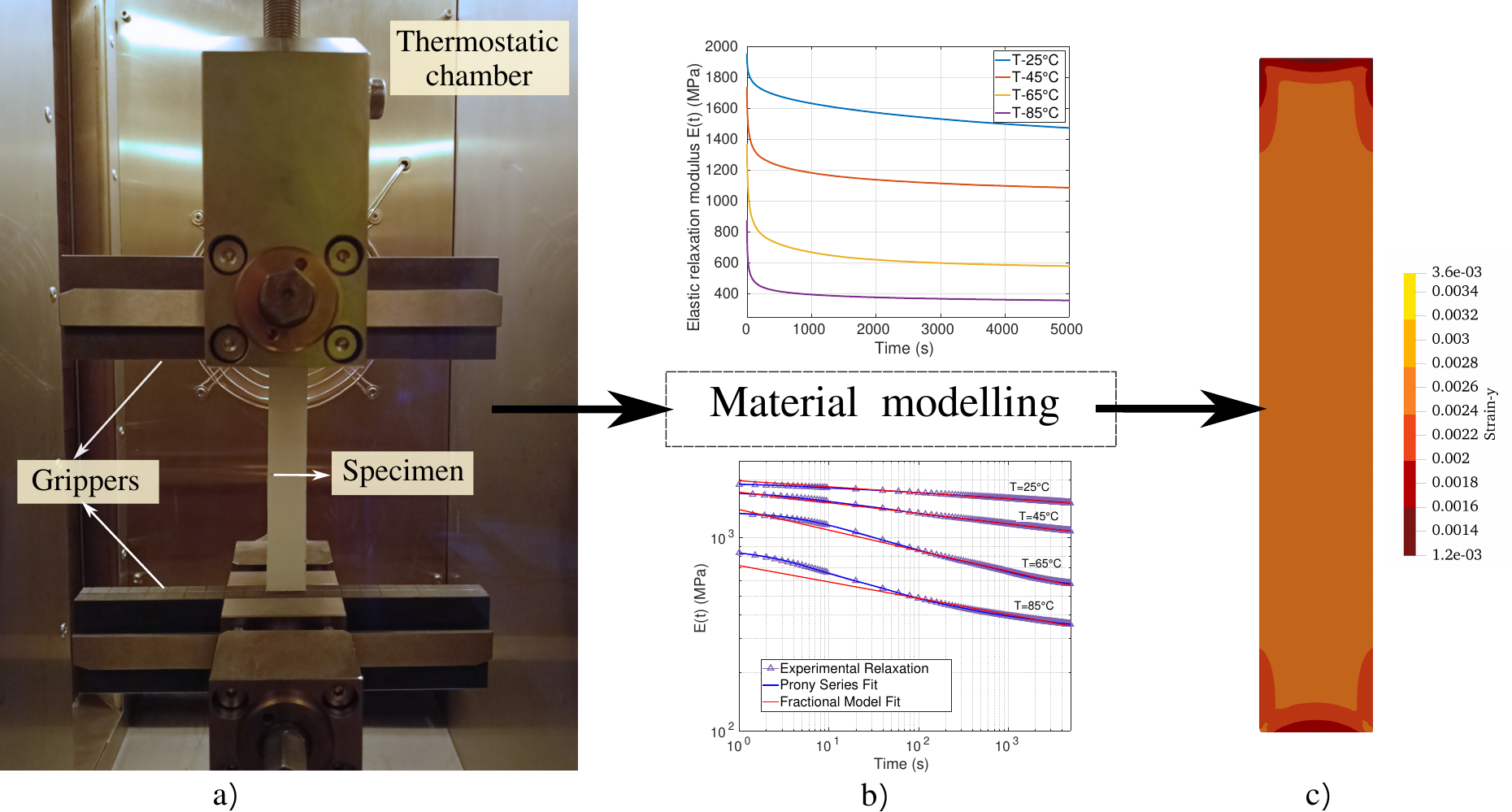}
    %\caption{A complete experimental setup with the loaded specimen highlighting important components: Create an interactive figure with procedure}
    \caption{ Schematic illustrating steps in modeling a) complete experimental setup with the loaded specimen highlighting important components b) Modelling response using prony series (PS) and fractional calculus-based models (FC), and c) FEM implementation for simulations.}
    \label{MachineNOmen}
\end{figure}

The experiments described in this section have been conducted in the MUSAM-Lab of the IMT School for Advanced Studies Lucca. 
All the specimens were tested with the Zwick/Roell Z010TH universal standard testing machine equipped with a 1 kN load cell and with the Zwick/Roell BW91272 thermostatic chamber to perform the experiments at different imposed temperatures.
The detail of the experimental setup is shown in Fig. \ref{MachineNOmen}.
All specimens have the same dimensions to have a better comparison. The thickness of each sample varies according to the respective manufacturer. The details of the material composition and thickness of the tested back sheet specimens are given in Tab. \ref{tab:Mcomp}. The backsheet material may have multilayer or single-layer. 
All backsheet materials under study have antireflective coatings on both sides for protection. 
In the present study, a rectangular-shaped sample specimen is prepared from a large sheet provided by the manufacturer, in which a length-to-width ratio $\approx$ 5:2 is maintained throughout the sample length.
Samples are cut out very carefully using a sharp cutter. 
Dimensions are chosen according to standard sample sizes specified in the literature \citep{carollo2019identification, eitner2011thermomechanics}. 
It was also ensured that there would not be any effects from grippers' (edge) effects due to the presence of enhanced areas at sample ends. 
A rectangular sample is prepared with dimensions 50 mm $\times$ 20 mm $\times$ thickness mm of respective backsheet material as given in Tab. \ref{tab:Mcomp}. 
In all experimental results presented, we use nominal stress (PK1), and the corresponding nominal strain corresponds to the change in deformation considered with respect to the initial length of the gauge section.

\begin{table}[h!]
\centering
\begin{tabular}{|c | c  |c |c|}
    \hline
	Commercial Name & Abbreviation  &   Material composition & 
	Thickness ($\mu$m)  \\
	\hline
 	Reflexolar OSBS  & OSBS  & White PP, polyamide, clear polyester& 320 \\
  	Madico TPE-HD & TPE-HD  & White PVF, clear polyester, EVA & 285 \\
	Medico Reflekt LEAN & Reflekt Lean & Fluoropolymer, polyester, 
	polyolefine & 269 \\	
        Lexan FR25A & Lexan  & Polycarbonate & 297 \\ 
	\hline	
\end{tabular}
\caption{Commercial names and composition of the testing materials} 
\label{tab:Mcomp}
\end{table}

\subsection{Uniaxial ensile tests}
\begin{figure}[h!]
    \centering
    \includegraphics[width=1.0\linewidth]{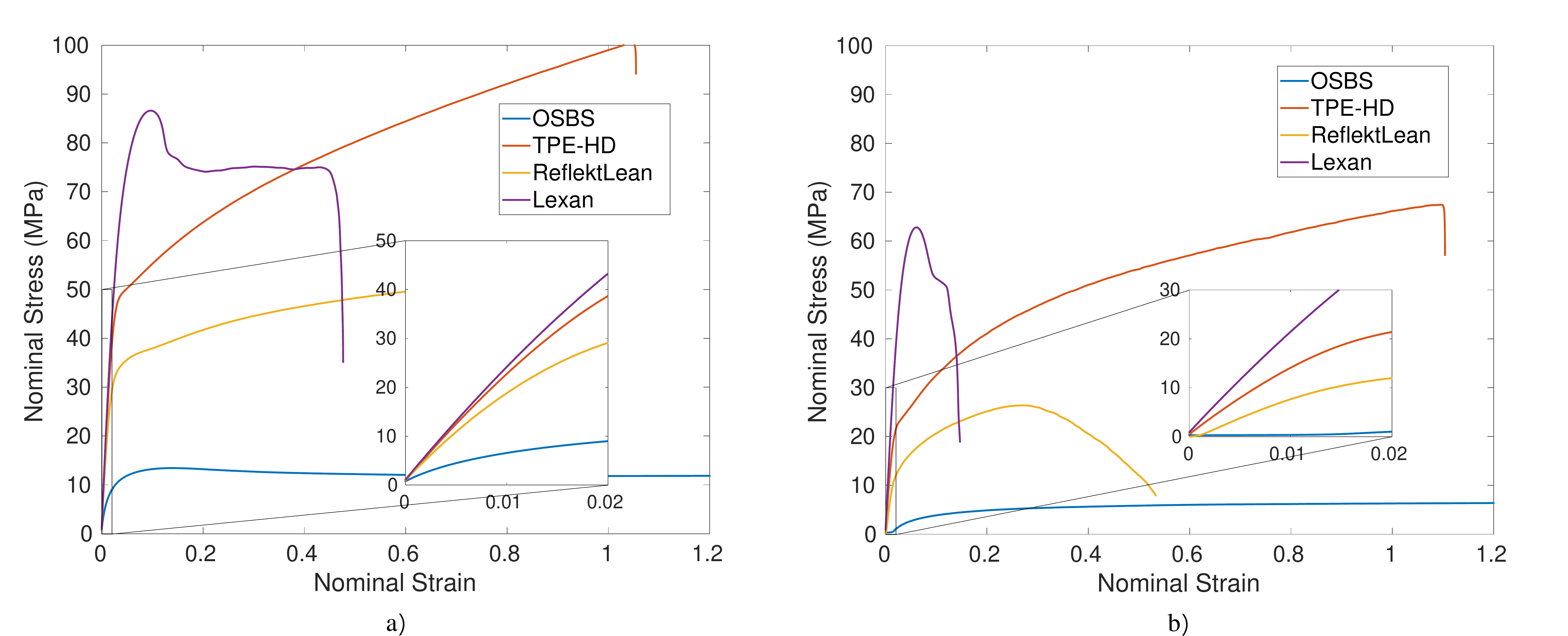}
    \caption{Tensile test performed until failure of specimens loaded with 0.1 mm/s at a) 25 \textdegree C and b) 85 \textdegree C respectively.}
    \label{fig:T25}
\end{figure}

The uniaxial tensile tests are often performed to determine the strength of materials. Material behavior can be categorized based on the relationship between stress, $\sigma$, vs. strain, $\epsilon$.
To understand the temperature effect on elasticity, the uniaxial tensile test was performed on the backsheets at an ambient $25$\textdegree C and at a high temperature of $85$\textdegree C, at which viscoelastic effects are more prominent and since some of the standard testing procedures for the assessment of the durability of PV modules requires high temperatures around $85$\textdegree C.
All backsheets were loaded by applying a constant strain over time.
Moreover, uniaxial tests up to failure were carried out with a constant loading rate of 0.1 mm/s at ambient and  $85$\textdegree C.
Plots of complete uniaxial tensile stress vs. strain are shown in Fig. \ref{fig:T25}. 
The backsheet materials, such as TPE-HD and ReflektLean, show plastic behavior before failure. The elastic modulus is calculated considering small strains up to 0.02 (2\%), which guarantees a linear response. The response becomes non-linear as strains start increasing further. 
A common observation is that the load-bearing capacity of backsheet materials has decreased at $85$\textdegree C compared to ambient temperature. As backsheets are made of polymers, the load required to straighten up the complex polymer chain at high temperatures decreases, which directly affects reducing their strength. 
The strength of Lexan was found to be the highest, irrespective of temperature. 
The strength of OSBS was found to be the lowest.
The response fitted linearly within the limit of 0.02 strain to find the elastic modulus under finite deformations. The bar chart compares elastic moduli for materials represented as shown in Fig.\ref{barTens}. The measured strength values for all the materials are collected in Tab. \ref{Ecompare}. 

\begin{minipage}{\textwidth}
  \begin{minipage}[b]{0.49\textwidth}
    \centering
    \includegraphics[width=1.1\linewidth]{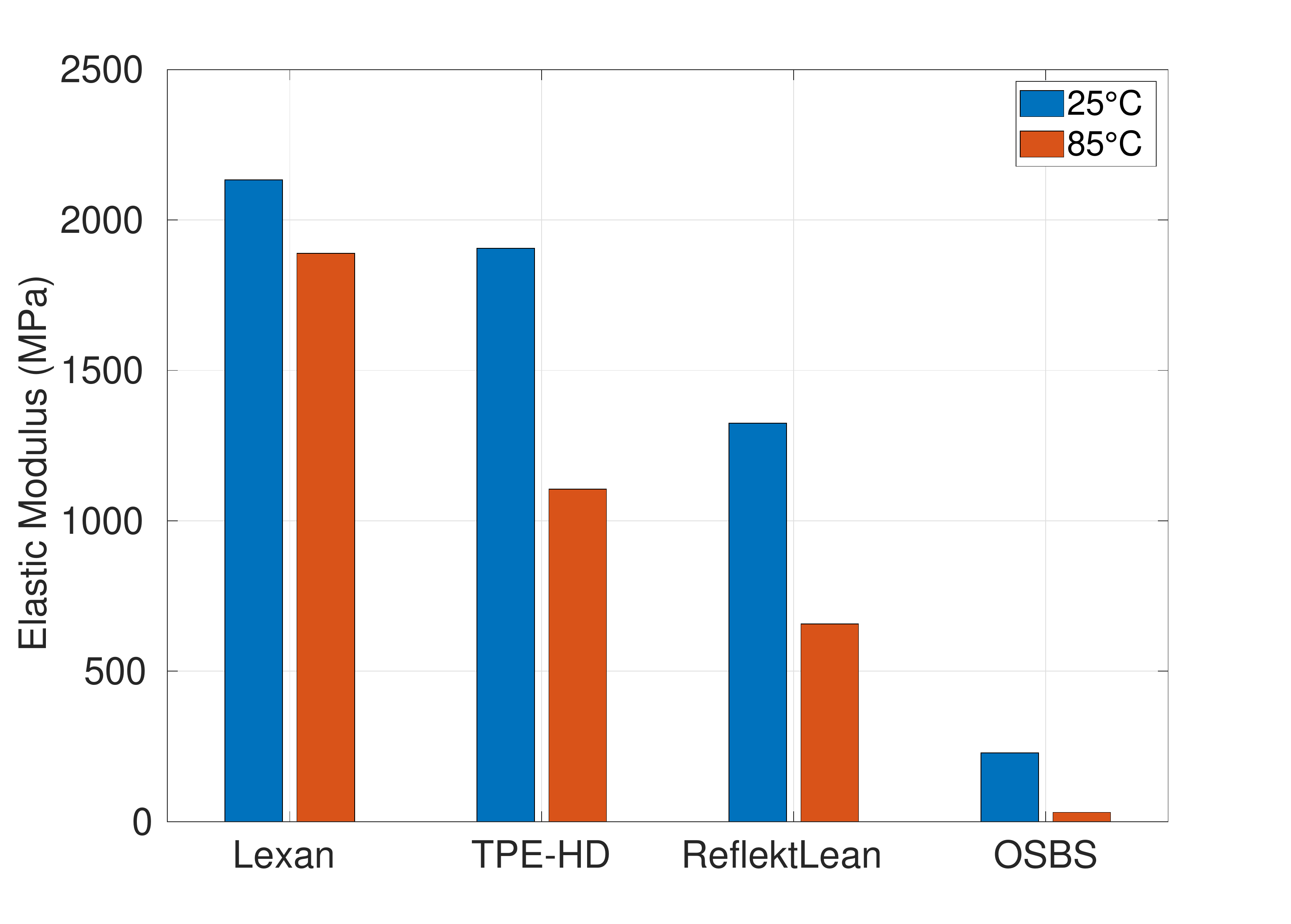}
    \captionof{figure}{Bar chart comparing elastic modulus of backsheets.}
    \label{barTens}
  \end{minipage}
  \hfill
  \begin{minipage}[b]{0.49\textwidth}
    \centering
    \begin{tabular}{cccc}
       \hline
       Material   & $\sigma_\text{25}$ (MPa)  &  $\sigma_\text{85}$ (MPa)  \\
       \hline
        %OSBS &  228.5 & 30.79 \\
        OSBS &  4.57 & 0.6158 \\
        %TPE-HD & 1906   & 1105 \\
        TPE-HD & 38.12   & 22.10 \\
        %ReflektLean &  1324 & 657.2\\
        ReflektLean &  26.48 & 13.14\\
        % Lexan& 2133  & 1889     \\
          Lexan& 42.66  & 37.78     \\
        \hline
      \end{tabular}
      \captionof{table}{Evaluated tensile strength of backsheet materials.}
      \vspace{60pt}
      \label{Ecompare}
    \end{minipage}
  \end{minipage}

\subsection{Relaxation tests}
\begin{figure}[h!]
    \centering
    \includegraphics[width=\linewidth]{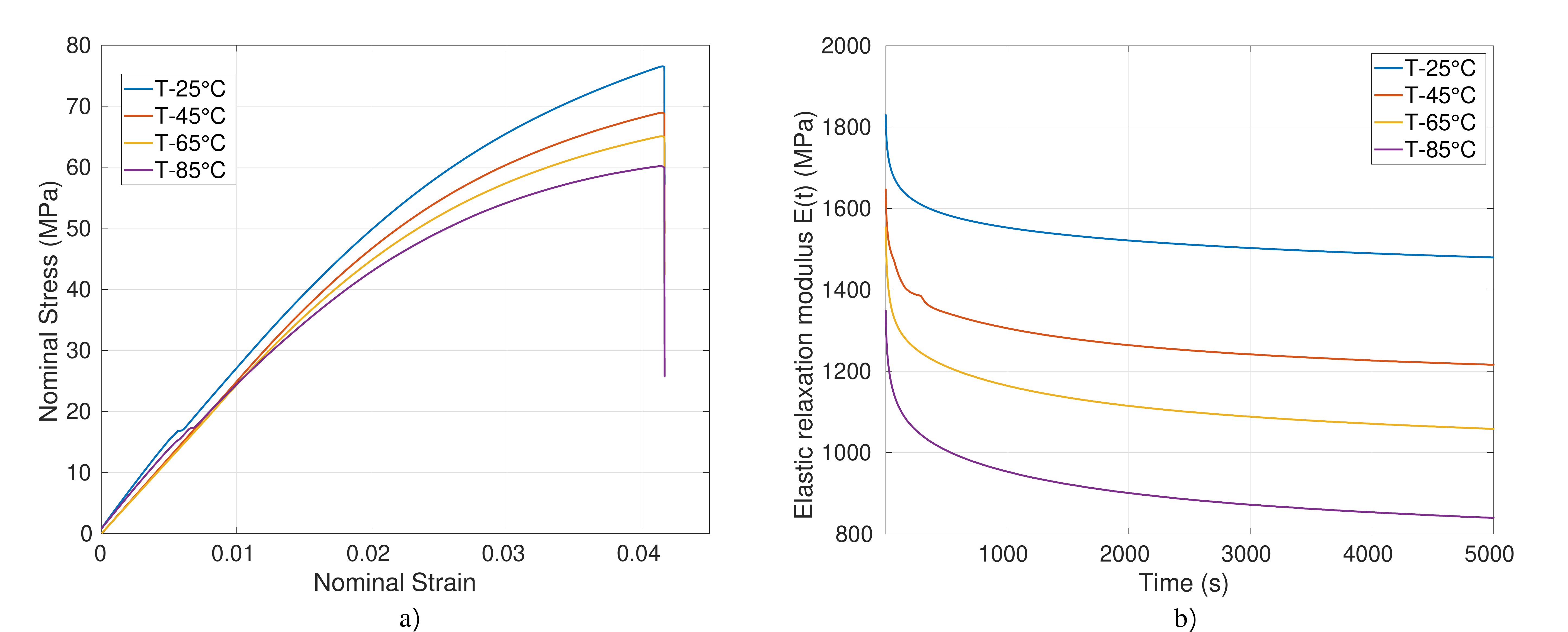}
    \caption{Temperature-dependent relaxation test response of Lexan back sheet: a) Nominal stress over nominal strain, and b) change in the elastic relaxation modulus over time (s).}
    \label{fig:LexanRelax}
\end{figure}

\begin{figure}[h!]
    \centering
    \includegraphics[width=\linewidth]{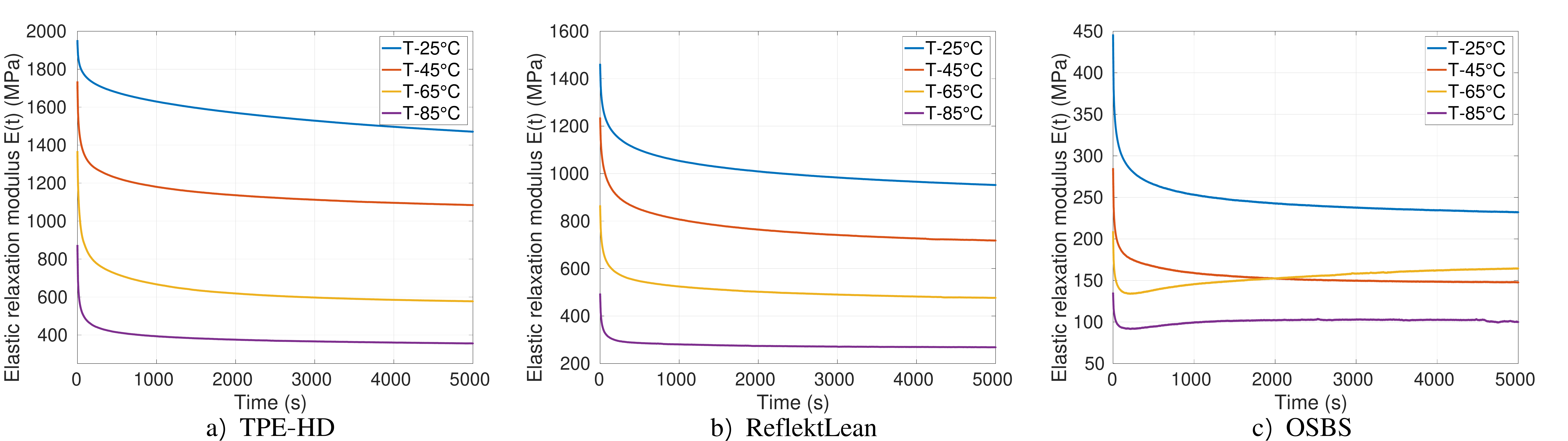}
    \caption{Temperature dependent relaxation test: response of change in the elastic relaxation modulus over time (s) for a) TPE-HD, b) Lean, and c) OSBS back sheet materials respectively.}
    \label{fig:Relax_all}
\end{figure}

Following the experimental analysis performed in \citep{eitner2011thermomechanics,borri2018fatigue}, at 85 \textdegree C, the backsheet material experiences thermal stress around 4-8 MPa which is sufficient enough to support viscoelastic straining/relaxation effects. 
Therefore, a series of relaxation tests were performed by holding the sample at corresponding constant strains for a long duration of time at a given temperature. 
In the current study, relaxation tests were carried out at constant strains but at different temperatures in the range 25-85 \textdegree C, to observe temperature-dependent effects. 
The initial loading rate was kept constant until reaching the corresponding constant strain, using 0.1 mm/sec for all cases. 
We have kept the initial loading rate constant so that results will be in line with uniaxial tests and valid comparisons can be carried out.
The instantaneous elastic modulus can be estimated by considering the problem as one-dimensional $E_{ins}= \frac{\sigma}{\epsilon_{0}}$. 
Corresponding values are collected in the Tab. \ref{Erelaxd}. 
It can be observed that values of the instantaneous elastic moduli are very much within the range of elastic moduli estimated previously from the tensile test. 
Fig. \ref{fig:LexanRelax} shows the relaxation response of the Lexan back sheet at corresponding temperatures. Instantaneous elastic modulus reduces as temperature increases. The relaxation response of other materials is represented in Fig. \ref{fig:Relax_all}.  All materials except OSBS follow the power-law type relaxation behavior at higher temperatures. As OSBS has the lowest elastic strength compared to others, it relaxes faster. 
A backsheet is a layered composition of different polymeric materials, every material has its glass transition $T_{g}$, which affects the relaxation response. 
Due to material composition, OSBS has lower $T_{g}$, affecting its relaxation response. 
As temperature increases beyond $T_{g}$, compared to other materials OSBS loses its power-law behavior.
These elastic relaxation responses of backsheet materials can be modeled into different material models. 


\begin{table}[h!]
    \centering
    \begin{tabular}{c c c c c c}
    \hline
     Material & Strain ($\varepsilon_{0}$)  &  $E_{25}$ (MPa) &  $E_{45}$ (MPa) &  $E_{65}$ (MPa) &  $E_{85}$ (MPa) \\
     \hline   
        OSBS   &  0.02 & 446.18 & 285.2  & 175.68 & 135.43\\    
      TPE-HD   & 0.02  & 1914.9 &  1735.8 &  1368.7   & 874.31\\
RefelektLean   & 0.02  & 1462.8 &  1236.7  &  866.47 & 495.03 \\
       Lexan   & 0.0417& 1836.6  &  1654.6 &  1561.8   & 1444.3\\
      \hline      
       \end{tabular}
    \caption{Instantaneous elastic modulus for the backsheet materials at the start of relaxation, for each corresponding test temperature.}
    \label{Erelaxd}
\end{table}

\begin{figure}[h!]
    \centering
    \includegraphics[width=\linewidth]{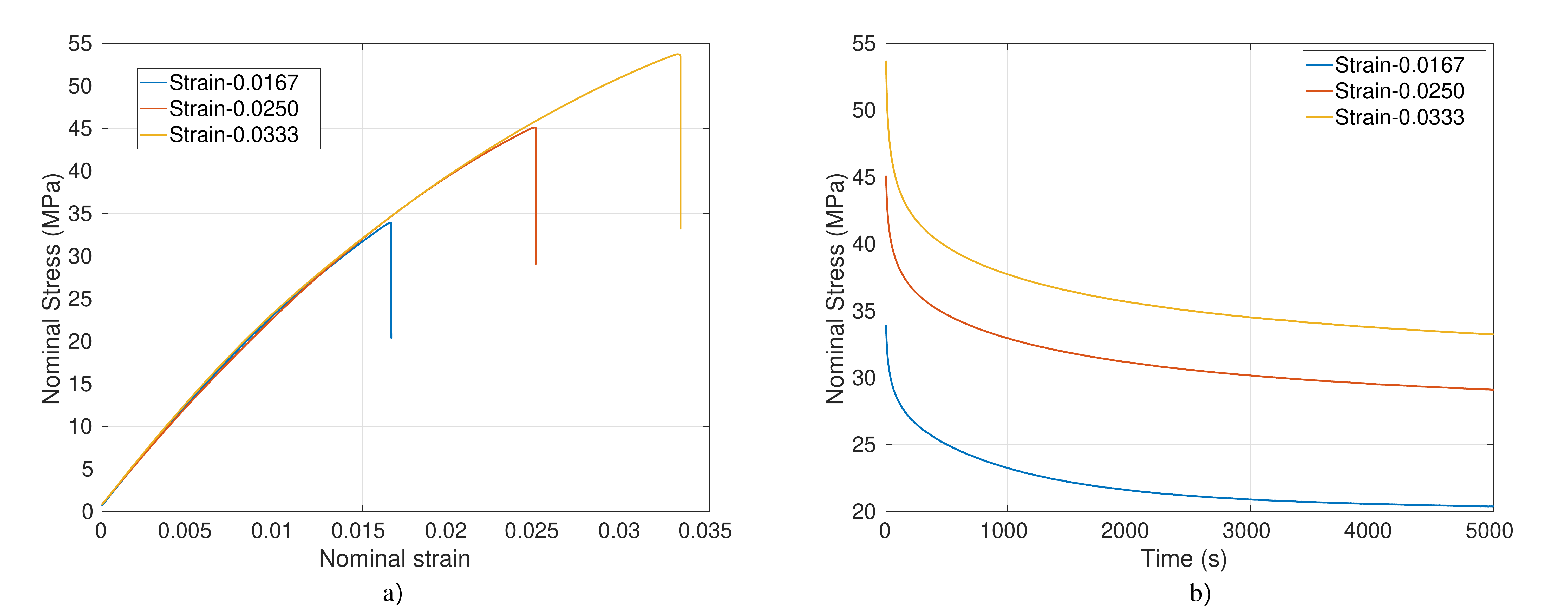}
    \caption{Relaxation test response of Lexan back sheet at constant temperature: a) Nominal stress over nominal strain, and b) change in the elastic relaxation modulus over time (s).}
    \label{fig:Lexan_const_relax}
\end{figure}

%\subsection{Strain effects}
% Comment on the instantaneous elastic modulus.\\
The relaxation tests were performed on Lexan specimens at a constant temperature for different strain levels to assess also how the maximum applied strain affects relaxation. Fig. \ref{fig:Lexan_const_relax} shows the relaxation response for different constant strains $\epsilon = 0.0167, 0.0250$, and 0.0333, for Lexan at 85 \textdegree C. Initially, stress is proportional to strain, but the response becomes nonlinear for higher strains. 
It is noted that at higher temperatures, specimens loaded at small strains relax faster than compared to what happens at larger strains. The study is performed to see changes in material relaxation parameters for different strains at the constant temperature which is discussed in Sec. \ref{Paraidentify}. 
%Uniaxial cyclic tests were also performed on all backsheet materials to check stress-hardening effects with increasing strains which may result in additional non-linear stress-strain response behavior, details of which are given in \ref{plasticity_data}.
%\subsection{Cyclic uniaxial tensile tests on back sheet materials: experimental results and interpretation}
%Comment on it during tensile tests only, parameters are not estimated. Which is not the focus of the current study. Put it into  the appendix section.

\section{Constitutive modeling of small-strain viscoelasticity} 
\label{Matmodels}
\begin{figure}[h]
	\centering
	\includegraphics[width=0.90\textwidth]{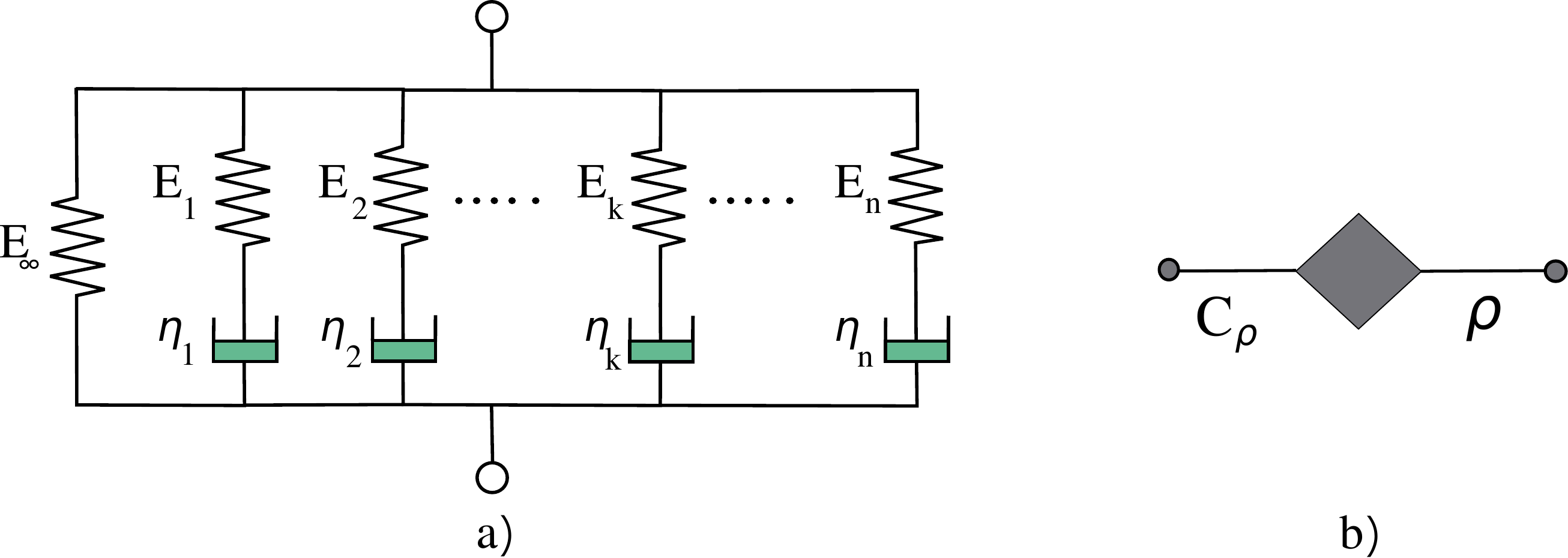}
	\caption{Schematic of representation  a) Generalized Maxwell Model b) Spring-pot Model for viscoelastic material models.}
	\label{GMM-schematic}
\end{figure}

Viscoelastic materials are characterized by their Relaxation and Creep functions $R(t)$ and $C(t)$, respectively. These functions describe the behavior of the material when a constant strain and constant stress are applied, respectively. Depending upon the choice of creep and relaxation response fitting with experimental tests, these functions can be modeled with the exponential response (PS model) or power law (FC model) as given below:
\begin{equation}
    R(t) = E_{\infty} +  \sum_{j = 1} ^{N} E_{j} e^{-\frac{t}{\tau_{j}}}
    \text{;} \hspace{40pt}
    C(t) = \frac{1}{E_{\infty}}  +  \sum_{j = 1} ^{N}  \frac{1}{E_{j}}  \left [ 1- e^{-\frac{t}{\tau_{j}}} \right] \hspace{10pt} %\text{Prony series model}
    \label{PSEq}
\end{equation}
\begin{equation}
    R(t) = \frac{A_{\alpha} t^{-\alpha}}{\Gamma(1- \alpha)}
    \text{;} \hspace{40pt}
    C(t) = \frac{t^{\alpha}}{A_{\alpha} \Gamma(1+\alpha)} \hspace{10pt} %\text{Fractional calculus model}
    \label{FCeq}
\end{equation}
where $E_{\infty}$ and $E_{i}$ are materials parameters for relaxation moduli corresponding to the fixed spring and the $i^{th}$ spring of Maxwell's arm in the PS model whereas,
$\Gamma(\cdot)$ is the Euler gamma function, $\alpha$ is a real number $0 \le \alpha \le 1$, and $A_{\alpha}$ is a material parameter. 
All these material parameters are evaluated by fitting creep or experimental relaxation curves. 
The constitutive models can be represented with physical means as shown in Fig. \ref{GMM-schematic}
In the linear viscoelasticity framework, the Boltzmann superposition principle allows us to obtain the response of a material when the imposed stress $s(t)$ or strain history $e(t)$ is not constant and can be expressed in two forms: 
\begin{equation}
         s(t) = \int_{0}^{t} R(t- \tau)\dot{e}(\tau)d\tau
         \label{strEQ}
\end{equation}
\begin{equation}
         e(t) = \int_{0}^{t} A(t- \tau)\dot{s}(\tau)d\tau
\end{equation}
These integrals are called hereditary integrals because the actual value of $s(t)$ (or $e(t)$) depends on the previous history of $e(t)$ (or $s(t)$).
Substitution of Eq.\ref{PSEq} and Eq.\ref{FCeq} into the above Eqs. gives the corresponding constitutive law.
Let's consider any solid body occupying region $\Omega   \subset  \mathbb{R}^{n_{\text{dim}}}$ $(n_{\text{dim}} = 1,2,3  \text{ spatial dimensions})$. 
Let \textbf{u} be the displacement field. Corresponding strain field can be defined as $\varepsilon := \text{sym} \left ( \nabla \textbf{u} \right)$. The external boundary $\partial \Omega  \subset 
\mathbb{R}^{n_{\text{ndim}}-1}$ composed of two open, disjoint regions, $ \partial \Omega = \overline{ \partial \Omega_{u}  \cup \partial \Omega_{t} }$ such that $u = 
\bar{u}$ on $\partial \Omega_{u}$ and $t= \bar{t}$ on $\partial \Omega_{t}$. For such a body, the total energy functional can be given as:
\begin{equation}
	\Pi  := \int_{\Omega} \Psi( \varepsilon) d\Omega - \int_{\Omega} 
	\textit{f}_{\textit{v}} \cdot {u} \hspace{2pt} d\Omega - \int_{\partial 
		\Omega} \bar{t} 
	\cdot {u} \hspace{2pt}d\partial \Omega 
	\label{visEq}
\end{equation}
% \textcolor{red}{why the $\cdot u$ in the integrals?}
Where $\textit{f}_{\textit{v}}$ is a body force and $ \Psi (\varepsilon)$ is elastic energy density function which is defined as $\Psi (\varepsilon) = \varepsilon(u)^{T}:\sigma(u)$, where $\sigma(u)$ is the stress which can also be given as $\sigma(u):= \mathbb{C}: \varepsilon(u)$ where $\mathbb{C}$ is a constitutive material tensor that is obtained through different material models.

\subsection{Generalised Maxwell model for linear Viscoelasticity}

The behavior of viscoelastic materials under loading is usually represented by conceptual models composed of elastic and viscous elements which correspond to  springs and dashpots.
A more general form of this is PS models, where an isolated spring is attached in parallel to several series spring-dashpot models, which are also called Maxwell's arms, to get a better viscoelastic response. 
The representative model is shown in Fig. \ref{GMM-schematic}.
As there are 2N+1 unknown parameters required to fit the experimental test, these models can give a good fit compared to other models.
However, it can be cumbersome since all the parameters should also have physical meaning, rather than just numerical parameters.
In most of the implementations, the most convenient choice is to use volumetric $K_{R}(t)$ and deviatoric relaxation $G_{R}(t)$ (or creep) functions \citep{taylor1970thermomechanical}. 
In the case of the PS model, these functions can be taken as a series of exponential responses similar to Eq. \ref{PSEq} as,
\begin{equation}
    G_{R}(t) = G_{\infty} +  \sum_{j = 1} ^{N} G_{j} e^{-\frac{t}{\tau^{g}_{j}}}; \hspace{20pt} K_{R}(t) = K_{\infty} +  \sum_{j = 1} ^{N} K_{j} e^{-\frac{t}{\tau^{k}_{j}}}
    \label{EqSTRrelax}
\end{equation}
Where, $G_{\infty},G_{i} $ and $ K_{\infty},K_{i} $ are corresponding shear and bulk relaxation modulii in fixed and $i^{th}$ maxwell's arms respectively 
with $\tau^{g}_{i}$ and $\tau^{k}_{i}$ are the corresponding relaxation time periods. 
Following \citep{holzapfel1996new}, bulk relaxation modulus can be taken as constant $K_{R}(t) = K^{v}$. 
This is because polymers show a predominant visco-elastic behavior in shear deformation rather than in volumetric expansion. It also limits the number of unknowns to evaluate.
Now, the strain-stress relationship can be given from Eq. \ref{strEQ} as:
\begin{equation}
   \bar{\sigma} (t) =  \int_{0}^{t} R(t- \tau)\dot{\varepsilon}(\tau)d\tau = K^{v} \theta + 2 \int_{0}^{t} G_{R}(t- \tau)\dot{e}(\tau)d\tau
   	\label{NetStress}
\end{equation}
where, $\varepsilon$ is actual strain can be given as, $\varepsilon = \textit{\textbf{e}} + \frac{1}{3} \theta$. In which, $\theta = tr(\varepsilon) \cdot I$  and $\textbf{e}$ 
are the corresponding second-order volumetric and deviatoric tensors. $K^{v}$ is the time-independent bulk modulus, and $\dot{e}(\tau)$ shows the rate of change of deviatoric strains with time.
Following Eq. \ref{EqSTRrelax}, $G_{\infty}$ is the shear modulus in the pure spring batch, which is also called the long-term shear modulus, and N denotes the number of Maxwell's arms. 
The term $\tau_{i} = \frac{\eta_{i}}{\mu_{i}}$, ($\mu_{i}$ is the shear modulus) is the relaxation time for each dashpot branch, which controls the time it takes for the stress to relax (higher the $\tau_{i}$ 
means the longer stress relaxation times). 
$\eta_{i}$ is the viscosity of the $i^{th}$ dashpot, which basically defines the strain rate dependence of material which can be taken as constant. 
The deviatoric stress $\bar{\sigma}_{dev}(t)$ from Eq. \ref{NetStress} can be written as:
\begin{equation}
	\begin{split}
		{\bar{\sigma}_{dev}}(t) &  = 2 G_{\infty} e(t) +  \sum_{i = 1}^N 2  
		\int_{0}^{t} G_{i} e^{-\frac{t-\tau}{\tau_{i}}} \dot{e}(\tau) d\tau
	\end{split}
	\label{feapsplit2}
\end{equation}
Using the time integration scheme for implementation of semi-analytical integration of the convolution integral, which is presented in \citep{taylor1970thermomechanical,londono2016prony}, 
the deviatoric stress ${\bar{\sigma}_{dev}}(t)$ at time $n+1$ for a time step of $\Delta t$ can be written in the form differential operators as:
\begin{equation}
	\begin{split}
		{\bar{\sigma}_{dev}}(t) &  = 2 \left [ G_{\infty} e(t) + \sum_{i=1}^{N} \left ( G_{i} \rm{exp} \left (-\frac{\Delta t}{\tau_{i}} \right ) h^{i}_{n+1}  \right ) \right ]
	\end{split}
	\label{feapsplit2}
\end{equation}
where $h^{i}_{n+1}$ is the stress corresponding to the $i^{th}$ Maxwell arm at current  time $n+1$ which can be given as an update integral considering strains at the current step and previous strains:
\begin{equation}
    h^{i}_{n+1} = \rm{exp} \left (-\frac{\Delta t}{\tau_{i}}  \right )h^{i}_{n} + h_{vis} \left ( e_{n+1} - e_{n} \right )
    	\label{Hmain}
\end{equation}
Where $h^{i}_{n}$ takes care of the previous time-step history of $h^{i}$, and $\rm{h_{vis}}$ is given by
\begin{equation}
    \rm{h_{vis}} = \frac{\tau_{i}}{\Delta t} \left ( 1- \rm{exp} \left( - \frac{\Delta t}{\tau_{i}} \right )  \right )
    \label{hdiff}
\end{equation}
Finally, the constitutive law for viscoelastic solids with moduli in the form of a Prony-series is obtained by directly substituting Eq.(\ref{hdiff}), (\ref{Hmain}) into Eq.(\ref{NetStress}) which defines the total stress $\bar{\sigma} (t)$ in the material. Due to the flexibility of accommodating a number of unknowns, in many cases of engineering interest, PS models can adequately reproduce the time-dependent features of viscoelastic materials \citep{knauss2015review,xu2018mathematical}

\subsection{Fractional calculus model for linear viscoelasticity}
The volumetric and the deviatoric creep/relaxation functions are both well-fitted by the pure power law function. Then their behavior can be reproduced by the spring pot model as shown in Fig. \ref{GMM-schematic}. In the case of the fractional calculus model, the substitution of relaxation function $R(t)$ into Eq. \ref{strEQ} leads to constitutive
laws that involve fractional operators, namely derivatives and integrals of real order \citep{samko1993fractional,alotta2018finite}. 
It is straightforward in which strain history is applied. Based on that, the corresponding stress history can be given as follows:
\begin{equation}
    s(t) = \frac{A_{\alpha}}{\Gamma(1-\alpha)}\int_{0}^{t}(t - \tau)^{-\alpha} \dot{e}(\tau) d\tau  = A_{\alpha} D_{t}^{\alpha}e(t)
\end{equation}
Where $D_{t}^{\alpha}$ represents the Caputo fractional derivative of order $\alpha$ \citep{podlubny1999fractional,scherer2011grunwald}, a convolution integral with a power law kernel.
It has been shown in Ref. \citep{paggi2015accurate,lenarda2022computational} that the behavior of the number of springs/dashpots in a classical viscoelasticity framework can be easily reproduced with a spring pot.
This is why using fractional viscoelasticity results in a significant reduction of mechanical parameters compared to using classical viscoelastic models. Similar to the PS model, volumetric $K_{R}(t)$ and deviatoric relaxation $G_{R}(t)$ (or creep) functions are chosen to represent viscoelastic behavior. 
In the case of the FC model, these functions can be taken as power law kernel similar to Eq. \ref{FCeq} as,
\begin{equation}
    G_{R}(t) = \frac{G_{\alpha} t^{-\alpha}}{\Gamma(1- \alpha)}; \hspace{20pt} K_{R}(t) = \frac{K_{\beta} t^{-\beta}}{\Gamma(1- \beta)}
    \label{FCSTRrelax}
\end{equation}
where $K_{\beta}$ and $G_{\alpha}$ are anomalous bulk and shear relaxation moduli, respectively, while $\beta$ and $\alpha$ are real numbers indicating the orders of bulk and shear power laws. Considering time-independent bulk modulus $K_{R}(t) = K^{v}$, Now, the strain-stress relationship can be given from Eq. \ref{strEQ} as:
\begin{equation}
   \bar{\sigma} (t) =  \int_{0}^{t} R(t- \tau)\dot{\varepsilon}(\tau)d\tau = K^{v} \theta + \frac{2 G_{\alpha} }{\Gamma (1-\alpha)}  \int_{0}^{t} (t- \tau) ^{-\alpha}\dot{e}(\tau)d\tau
   	\label{NetStress2}
\end{equation}
 The time-dependent shear relaxation modulus $G(t)$ is found via elastic/visco-elastic correspondence principle and inverse Laplace transform using the Mittag-Leffler special functions as given in \citep{alotta2018finite,bonfanti2020fractional}. The deviatoric stress tensor $ \bar{\sigma}_{dev}$ can be written with shear relaxation function $G_{\alpha}$ and $\alpha$ as follows:
\begin{equation}
	\bar{\sigma}_{dev}(t) = 2 \left [  \frac{G_{\alpha}}{\Gamma(1-\alpha)} 
	\int_{0}^{t}(t - \tau)^{-\alpha} \dot{e}(\tau) d\tau \right ] 
	= 2 G_{\alpha} D^{ \alpha}_{t} e(t)
\end{equation}
The Caputo fractional derivative of any function $f(t)$ can be given by using the Grunwald-Letnikov (GL) fractional derivative \citep{wei2001fe,schmidt2002finite}. The Grunwald-Letnikov approximation of the fractional derivative at the current time steps $n+1$ for a time step of $\Delta t$ , $D^{\alpha} f(t^{n+1})$ for $f$ function \citep{samko1993fractional} given as:
	\begin{equation}
	D^{\alpha} {f}(t) = (\Delta t)^{-\alpha} \sum^n_{j=0} 
	c_{j+1}(\alpha) f^{n+1-j}
	= (\Delta t)^{-\alpha} \left[ 
	f^{1} |  \dots  |  f^{n+1} \right]  \begin{pmatrix}    
	c_{n+1}(\alpha) \\
		\vdots \\
		c_1(\alpha)
	\end{pmatrix}
	\end{equation}
where the coefficients $c_{j}(\alpha)$ are defined by the recursive formula:
\begin{equation}
	c_{j}( \alpha ) = \left \{ \begin{array}{rcl}
		\frac{(j-1- \alpha)}{j}  c_{j-1}(\alpha) & \mbox{if}   & j> 1 \\
		\\
		1   & \mbox{if}  & j = 1
	\end{array}
	\right.
\end{equation}
Following this, the deviatoric stress ${\bar{\sigma}_{dev}}(t)$ at time $n+1$ for a time step of $\Delta t$ can be written in the form differential operators as:
\begin{equation}
	\bar{\sigma}_{dev}(t) = 2 G_{\alpha} D^{ \alpha}_{t} e(t) = 2 G_{\alpha} (\Delta t)^{-\alpha} \sum^n_{j=0} 
	c_{j+1}({\alpha}) e^{n+1-j}
\end{equation}
Coefficients are such that $c_{j}(\alpha) < c_{j+1}(\alpha) <0  $ for $j >1$  and $ \text{lim}_{j \rightarrow +\infty}  c_{j}(\alpha) = 0$. 
In the FEM  discretization, $f$ matrix stores the values of deviatoric strains $e$ of previous time steps into columns $\left[ \text{e}^{1} |  \dots  |  \text{e}^{n+1} \right] $ with increasing order respectively which acts as the memory of material. 
Using Eq. \ref{NetStress2} with the volumetric and the deviatoric creep/relaxation functions are both well fitted by pure power law function. Then their behavior can be reproduced by the spring pot model.
In many cases of engineering cases, spring pots may not adequately reproduce the time-dependent features of viscoelastic materials completely, then other fractional viscoelastic material models with different combinations of springs along with spring pots are used \citep{alotta2018finite}.

\subsection{Numerical implementation}
Viscoelastic constitutive laws are time-dependent equations that can be solved by the finite element method (FEM) through step-by-step numerical integration schemes.
We have implemented models into an implicit(Newton-Raphson) scheme; all of the components of stress at the end of a previous time step must be provided.
Here, we show details of the implementation of PS and FC models in a user element routine(UEL) into the finite element analysis program (FEAP), \citep{taylor2014feap}.
For the implementation of viscoelasticity in an implicit integration scheme, all of the components of stress and the jacobian at the end of a time step must be provided for each Gauss point.
The deviatoric and hydrostatic parts of the stress tensor are given by:
\begin{equation}
    {\bar{\sigma}_{dev}}(t)  = 2 \left [ G_{\infty} e(t) + \sum_{i=1}^{N} \left ( G_{i} \rm{exp} \left (-\frac{\Delta t}{\tau_{i}} \right ) h^{i}_{n+1}  \right ) \right ] = 2 G_{\alpha} D^{ \alpha}_{t} e(t)
    \text{;} \hspace{20pt} {\bar{\sigma}_{vol}}(t) = K^{v}\theta
\end{equation}
The weak form corresponds to the energy functional Eq. \ref{visEq} is derived by multiplying it by a virtual displacement v and integrating the result on the domain $\Omega$. Applying the divergence theorem:
\begin{equation}
    \int_{\Omega}  {\bar{\sigma}_{dev}} e(u): \varepsilon(\text{v}) d\Omega + \int_{\Omega} K^{v} \rm {div(u)} \cdot \rm{div(v)} = \int_{\Omega} 
	\textit{f}_{v} \cdot {v} \hspace{2pt} d\Omega + \int_{\partial 
		\Omega} \bar{t} 
	\cdot {v} \hspace{2pt}d\partial \Omega 
\end{equation}
Following discretization of $\Omega$ into elements such that $\Omega \rightarrow \Omega_{e}$. The trial solution functions space for displacement $v$ can be defined as,
\begin{equation}
    \mathcal{U}_{h}(u) = \left \{ u \in H^{1} (\Omega)| \nabla u \in L^{2} (\Omega); u =u_{d} \hspace{0.05in} on \hspace{0.05in}  \partial \Omega_{d}       \right \}
\end{equation}
Respective test functions space for displacement $v$ can be given as,
\begin{equation}
    \mathcal{V}_{h}(v) = \left \{v \in H^{1} (\Omega)| \nabla \delta u \in L^{2} (\Omega); v= 0 \hspace{0.05in} on \hspace{0.05in}  \partial \Omega_{d}       \right \}
\end{equation}
For the single element, displacement and it's gradients are approximated by Galerkin's method as follows,
\begin{equation}
    \begin{array}{cccc}
        u_{e} \approx N_{u} d_{u} & \nabla u_{e} \approx B_{u} d_{u}  &  v_{e} \approx N_{u} d_{v} & \nabla v_{e} \approx B_{u} d_{v}  
    \end{array} 
\end{equation}

Where variable $d_{u}$ are vectors of nodal values of displacements, in our implementation, the standard bi-linear shape functions $N$ are used for both displacements and phase field variables. 
For simplicity, the gradient of shape function $\nabla N$ is taken as $B$.
Through the insertion of the previous interpolation scheme for the displacement and phase field variable, the discrete versions of element residual vectors for both fields are given by,
\begin{equation}
    R_{u} = \int_{\Omega _{e}} \nabla v^{T} { \left ( \bar{\sigma}_{dev} + \bar{\sigma}_{vol} \right)} d\Omega - \int_{\Omega_{e}} v^{T} f d \Omega 
            -\int_{\partial \Omega_{e}} v^{T} \Bar{t} d \partial \Omega
\end{equation}
To obtain the solutions for which $R_{u} = 0$ and because the corresponding residuals are nonlinear, an incremental-iterative scheme using the Newton-Raphson method is employed,
\begin{equation}
     (d_{u})_{t+\Delta t} = (d_{u})_{t} - \left[ K^{u} \right]^{-1} (R_{u})_{t}
\end{equation}
where coefficients of $K^{u}$ can be given as follows,
\begin{equation}
    K^{u} = \frac{ \partial R_{u}}{ \partial d_{u} } = \int_{\Omega} B_{v}^{T} \bar{\mathbb{C}}^{vis} B_{u} d\Omega
\end{equation}
where $\bar{\mathbb{C}}^{vis}$ viscoelastic material tensor can be defined as,
\begin{equation}
\label{CvisUNdamaged}
   \bar{\mathbb{C}}^{vis} = \frac{\partial \Delta \sigma_{ij}^{n+1}}{\partial \Delta \varepsilon_{kh}^{n+1}} =
    \left( K^{v} -\frac{2}{3}G_{vis} \right) \delta_{ij} \delta_{kh} + G_{vis} \left( \delta_{ik} \delta_{jh} + \delta_{ih} \delta_{jk} \right)
\end{equation}
The components of the material tensor can be evaluated as follows:
\begin{equation}
    \frac{\partial \Delta \sigma_{ii}^{n+1}}{\partial \Delta \varepsilon_{ii}^{n+1}} = K^{v} + \frac{4}{3} G_{vis} \text{;   \hspace{20pt}$i= 1,2,3$ }\\
\end{equation}

\begin{equation}
    \frac{\partial \Delta \sigma_{ii}^{n+1}}{\partial \Delta \varepsilon_{jj}^{n+1}} = K^{v} - \frac{2}{3} G_{vis} \text{;   \hspace{20pt}$i,j= 1,2,3$;  $i \neq j$ }\\
\end{equation}

\begin{equation}
    \frac{\partial \Delta \sigma_{ij}^{n+1}}{\partial \Delta \varepsilon_{ij}^{n+1}} =  G_{vis} \text{;   \hspace{20pt}$i,j= 1,2,3$, $i \neq j$ }\\
\end{equation}
Where $K^{v}$ is the time-independent bulk modulus and  $G_{vis} =  G_{\infty}+ \sum_{i=1}^{N} G_{i} \rm{h_{vis}}$ and $G_{vis} =  G_{\alpha} \Delta t^{-\alpha}$ respectively considered for PS and FC models, which is only modification from a linear elastic material is the substitution of the factor. $\delta$ is Kronecker delta. It can be noted that for both models, Jacobian depends only on the value of $\Delta t$ and on the mechanical parameters. Fixed time-step increment $\Delta t$ is considered to obtain convergence results.
As compared to the prony series model, which only requires the history of stress at the previous time step, the fractional model needs complete history data of stress at Gauss points to calculate the stress at the current time step, which takes a significant amount of memory data. This is one of the main reasons why the FC Model takes a more extensive simulation time than the PS model. 
In order to reduce the amount of memory, one can use a larger time increment $\Delta t$ or truncate the memory of the material for the GL derivative. 
In this study, we have chosen optimum value for $\Delta t$ to improve computational effectiveness.

\section{Parameter identification procedure}
\label{Paraidentify}
Since the models for the viscoelastic material characterization were established in the previous section, here we describe the material parameter identification. 
There are various techniques suggested in the literature \citep{chen2000determining,suchocki2013determination} to identify the model parameters.
A more suitable choice is to use a relaxation response to get material parameters.
For the relaxation test, a constitutive relation for the period of constant strain can be given as $ \sigma(t) = E(t) \epsilon_{0}$,
where $E(t)$, is the relaxation function and $\epsilon_{0}$ is the strain at a time, $t=0$. 
Considering the material functions based on both the models can be given by expressions discussed previously as :
\begin{equation}
    E(t) =  A_{\alpha} t^{-\alpha}/ \Gamma(1-\alpha), \hspace{25pt} 0 \le \alpha \le 1   \hspace{30 pt} \text{Fractional calculus Model}
\end{equation}
\begin{equation}
    E(t) =   E_{\infty} + \sum_{j = 1} ^{N} E_{j} e^{-\frac{t}{\tau_{j}}}       \hspace{90 pt} \text{PS Model}
\end{equation}
where $R(t)$ is already available from experimental relaxation tests as given in Fig. \ref{fig:LexanRelax}.
Consider vector $p$, a set of model parameters $p = (p_{1}, p_{2},.....,p_{m})$ which collects unknown coefficients of the Prony series, or $(A, \alpha)$ for the Fractional model. Vector $p$ take values in parameter space $P$, containing all the admissible values for $p$. 
This feasible set $P$ also takes into account the restriction of model parameters. The problem of estimating the coefficients is addressed as an optimization problem which has the objective of minimizing cost function $\phi(p)$ defined as:
\begin{equation}
    \phi(p) = \sum_{i= 1}^{N} \left ( f_{i}(p) \right )^{2},  \hspace{20 pt} \text{for}  \hspace{20 pt}  f_{i} \left ( p \right ) = E_{i}^{\text{exp}} - E_{i}^{\text{model}}(p)
\end{equation} 

\begin{figure}[h!]
	\centering
	\includegraphics[width=0.65\linewidth]{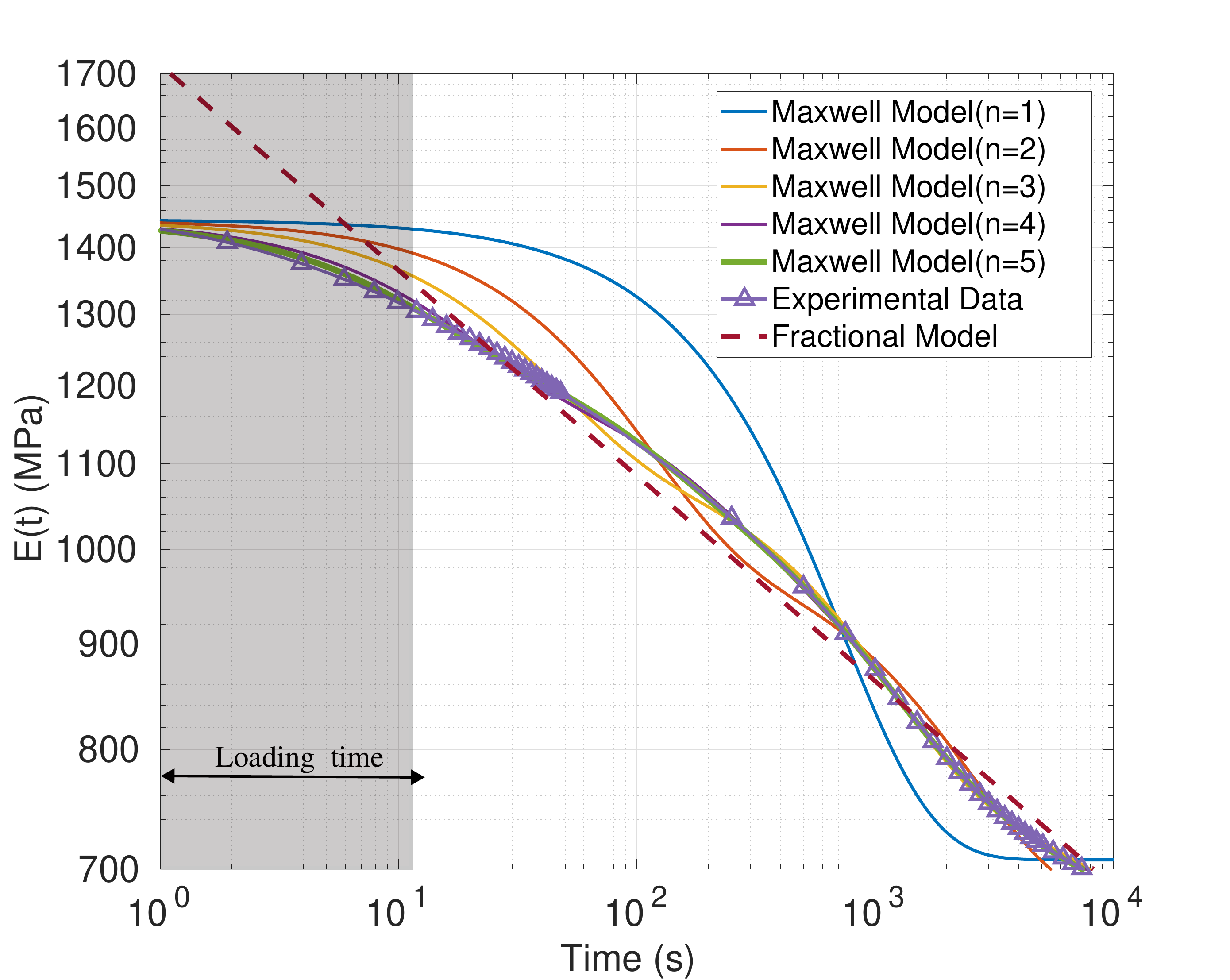}
	\caption{Parameter estimation for relaxation response of Lexan at 85 \textdegree C for Maxwell and Fractional Model.}
	\label{CurveFit}
\end{figure}

To minimize the cost function, $\phi(p)$ there are various algorithms reported in the literature, such as \citep{carollo2019identification} uses gradient descent algorithm and Particle Swarm Optimization (PSO) to get model parameters. In \citep{kraus2017parameter}, GUSTL was used to identify the Prony series from DMTA data, while parameter estimation based on Genetic Algorithms (GA) was done  in \citep{kohandel2008estimation}. 
It was found that the Nonlinear Least Squares method based on the Levenberg-Marquardt optimization algorithm solves the minimization problem effectively \citep{ranganathan2004levenberg}.
Based on that, an interactive MATLAB program has been implemented to extract the model parameters from the present experimental relaxation tests.
In the case of the Prony Series (PS) model, there are $2n+1$ unknowns corresponding to $E_{\infty}$, $E_{i}$'s, and $\tau_{i}$'s. As the number of Maxwell's arms increases so do the unknowns. In literature, it was suggested that the values for $\tau_{i}$ can be chosen as multiples of decades \citep{paggi2015accurate,eitner2011thermomechanics}. One of the reasons to choose multiple decades is to accommodate a complete relaxation response. 
For example $\tau = ( 10^{0}, 10^{2}, 10^{4})$. So that PS model parameters would reduce to $n+1$. The idea behind this is that every Maxwell's arm has a local impact on the shape of $E(t)$. Parameters 
$\tau_{i}$ define the range of the impact within time $t$ and $E_{i}$ sets the height of the model curve. 
As the problem gets constrained in the time range, a higher number of Maxwell's arms is needed to get an accurate response. 
Instead of fixing the time range $\tau_{i}$'s, the current study focuses on finding the optimum time $\tau_{i}$ for dashpots which can minimize the cost function $\phi(p)$. Though the total unknown model parameters are $2n+1$, the overall number of parameters to find gets reduced compared to fixing the time range. It also guarantees physical meaning to the outcome of the identification problem.
Fig. \ref{CurveFit} shows the plot of $E(t)$ fitted with up to 5 Maxwell's Arms for a relaxation response for Lexan at 85 \textdegree C. 
For $n=1 $ to 3 Maxwell's arms, the response of $E(t)$ is wavier, but as the number $n$ of arms increases, the error between the experiment and model decreases as shown in Tab. \ref{ErrMax}.
Using the same algorithm, fractional calculus (FC)  model parameters A and $\alpha$ are estimated for the relaxation response of Lexan, as shown in Fig. \ref{CurveFit}. 
The response of the Fractional model (FM) is linear on the double logarithmic scale with an objective function, $y = log $ \text{E}(t), having a slope of $m = -\alpha$ with a constant of $c = log \frac{A}{\Gamma (1- \alpha)}$. 
The initial value of the relaxation modulus depends on the constant. Since $\alpha$ is in line with the slope of relaxation, the corresponding values obtained from the Fractional model are higher than that of experimental data. But after a short duration, the Fractional Model gives a very good fit compared to other Maxwell models with many arms.
\\
The short-term period can be considered a loading time period most of the time \citep{glaesener2021viscoelastic}, and it is time to obtain the constant state of stress, $\sigma_{0}$ for creep or a constant state of strain $\epsilon_{0}$ as for relaxation experiments. Though the Fractional calculus model is not able to predict short-time relaxation behavior, these models are very effective for long-term response \citep{bonfanti2020fractional,paggi2015accurate}. Most modeling problems are done for a long-term response which accounts for the viscoelastic relaxation. These models are very effective since they require only 2 unknowns to describe the complete relaxation as compared to $2n+1$ for models based on the Prony series. All experimental data with MATLAB code for estimating model parameters with optimization is provided in \ref{SupplementaryDAT}, the supplementary data of this article.

\begin{table}[h!]
    \centering
    \begin{tabular}{c|c cc c c c|c}
    \hline
     Material Model  & PS  &1 & 2 & 3 & 4 & 5 & FC\\
     \hline
     RSME Error $(\%)$ & &123.320  & 12.896 & 0.887  &0.101 & 0.014 & 1.301 \\
     \hline
    \end{tabular}
    
    \caption{Error analysis with increasing Maxwell's arm and Fractional Calculus Model.}
    \label{ErrMax}
\end{table}

\begin{figure}[h!]
    \centering
    \includegraphics[width=1.0\linewidth]{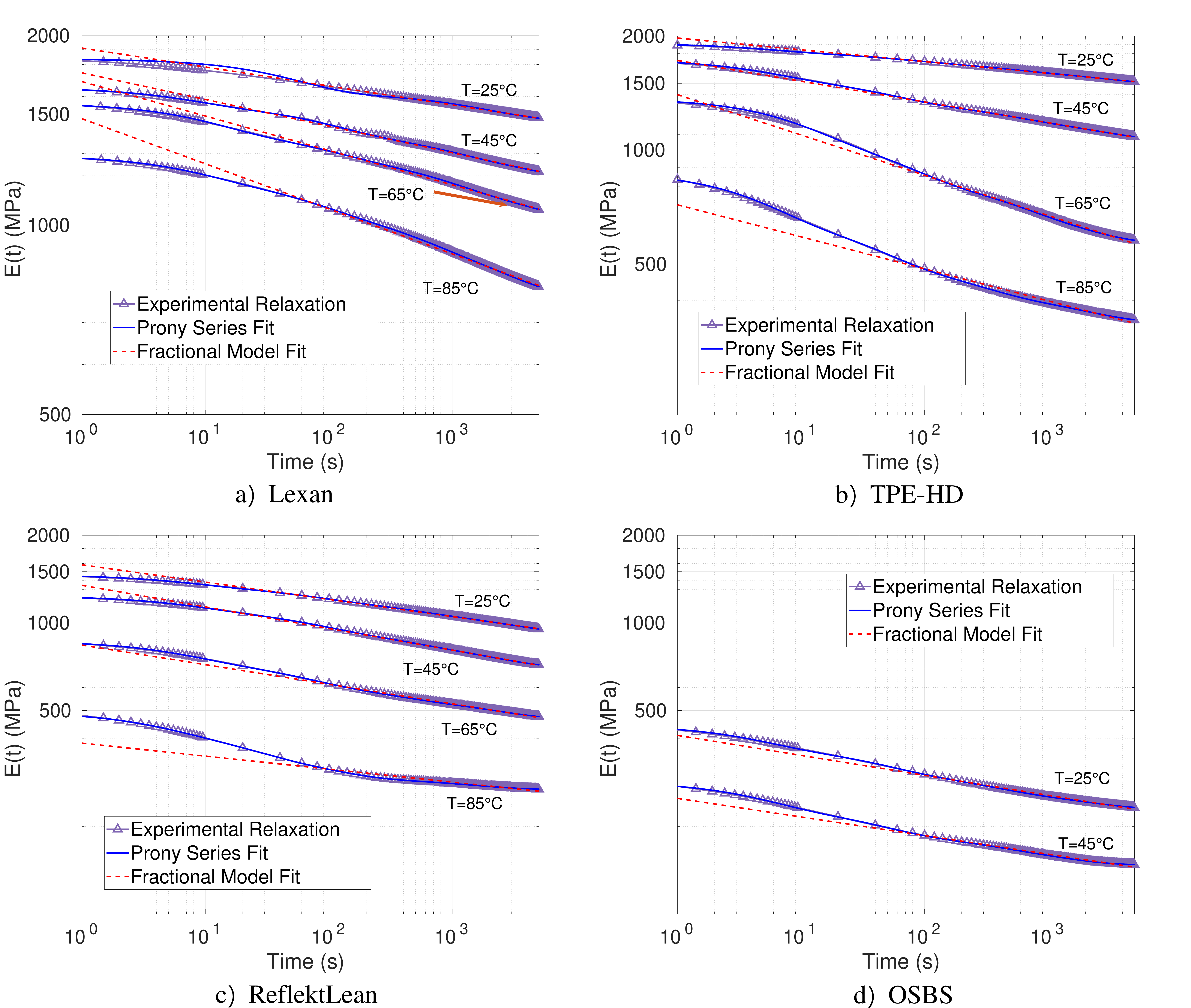}
    \caption{Experimental results with numerical fitting for elastic relaxation modulus $E(t)$ versus time response for a) Lean and, b) OSBS back sheet materials at different temperatures.}
    \label{fig:Model-Lean/OSBS}
\end{figure}

\subsection{Parameter estimation for temperature-dependent relaxation response of backsheet materials}
The relaxation response of backsheet materials at different temperatures for constant strains can be modeled using PS and FC models, as shown in Fig. \ref{fig:Model-Lean/OSBS}.  
The model parameters for both material models are obtained for all backsheet materials based on the relaxation response obtained from experiments as shown in Tab. \ref{PSmodel} and \ref{FMmodel}, respectively. 
An optimum number of Maxwell's arms was identified, which was used to find parameters. As the number of PS parameters increases beyond 4-5 terms, finding the physical meaning of terms is difficult. 
But valid observations can be made from fractional calculus parameters. As temperature increases, slope $\alpha$ increases until temperature reaches $T_{g}$, which is the glass transition temperature at which constituent polymer material loses its crystalline phase. A similar analysis for EVA is reported in \citep{paggi2015accurate,eitner2011thermomechanics}. 
As backsheet materials are layered composites, finding the exact transition temperature is difficult. $T_{g}$ is helpful in finding Williams- Landel-Ferry parameters (WLF) theoretically expected shift function) which plays an important role in Time-Temperature-Superposition-Principle (TTSP) to get an approximate response by shifting the master curve. Here we can estimate the approximate $\alpha$ from its variation corresponding to the temperature. Except for OSBS, for all other backsheet materials, $\alpha$ increases as temperature increases till higher temperatures around 65 \textdegree C, but for OSBS, it starts decreasing from room temperature. One of the main reasons for having lesser $\alpha$ for OSBS is its lower $T_{g}$. Refer \citep{pander2011thermo, eitner2011thermomechanics} to understand the role of $T_{g}$ on the viscoelastic materials. Every time, it isn't easy to get a master curve for viscoelastic materials such as backsheet, so performing temperature-dependent tests with characterization gives an idea about the change of material parameters with temperature. 
\begin{table}[h!]
    \centering
    \begin{tabular}{c c c c c c}
    \hline
     Material & Temperature & Strain &PS  & Parameters Identified  & RSME \\
       & (\textdegree C )& $\varepsilon$ & Terms& $E_{i}$ (MPa), $\tau_{i}$ (s)  & Error $(\%)$  \\
     \hline   
           &   &  &  & $E_{\infty} = 229.47 $, $E_{1} = 71.53 $, $ \tau_{1} = 4.59 $  & \\
           & 25  &  0.02  & 4  & $E_{2} = 64.54 $, $\tau_{2}= 42.60 $, $ E_{3} = 44.051 $  & 1.58\\
           &     &      &   & $\tau_{3} = 288.602 $, $ E_{4}= 36.59 $, $ \tau_{4} = 1980.18 $  & \\
           \cline{2-6}
    OSBS   &  &  &  & $E_{\infty} = 147.48 $, $E_{1} = 51.264 $, $ \tau_{1} = 4.85 $  & \\
           &  45    & 0.02      & 4  & $E_{2} = 44.41 $, $\tau_{2}= 39.26 $, $ E_{3} = 16.66 $  & 1.01\\
           &     &      &   & $\tau_{3} = 267.65 $, $ E_{4}= 25.38 $, $ \tau_{4} = 1201.95 $  & \\        
      \hline
           &   &  &  & $E_{\infty} = 1490.58 $, $E_{1} = 99.523 $, $ \tau_{1} = 5.431 $  & \\
           & 25    &   0.02   &  4 & $E_{2} = 92.24 $, $\tau_{2}=57.21 $, $ E_{3} = 91.41 $  & 2.23\\
           &     &      &   & $\tau_{3} = 373.25 $, $ E_{4}= 141.17 $, $ \tau_{4} = 2881.27 $  & \\  
           \cline{2-6}
           &   &  & & $E_{\infty} = 1058.275 $, $E_{1} = 106.69 $, $ \tau_{1} = 3.10 $  & \\
           &  45   &  0.02    & 5   & $E_{2} = 134.925 $, $\tau_{2}= 15.96 $, $ E_{3} = 163.601 $  & 0.125\\
           &     &      &   & $\tau_{3} = 67.19 $, $ E_{4}= 116.425989 $, $ \tau_{4} = 447.94 $  & \\   
           &     &      &   & $ E_{5}=155.86 $, $ \tau_{5} = 2813.216 $  & \\           
           \cline{2-6}
   TPE-HD  &   &  &  & $E_{\infty} = 567.25 $, $E_{1} = 268.456 $, $ \tau_{1} = 10.4 $  & \\
           & 65    &   0.02   & 4  & $E_{2} = 258.15 $, $\tau_{2}=64.9 $, $ E_{3} = 136.014 $  & 2.06\\
           &     &      &   & $\tau_{3} = 463.862 $, $ E_{4}= 138.82 $, $ \tau_{4} = 1948.052 $  & \\
          \cline{2-6}
           &   &  &  & $E_{\infty} = 351.01 $, $E_{1} = 211.414 $, $ \tau_{1} = 5.13 $  & \\
           &  85   &  0.02    & 4   & $E_{2} = 156.97 $, $\tau_{2}=38.81 $, $ E_{3} = 87.39 $  & 1.87\\
           &     &      &   & $\tau_{3} =238.517 $, $ E_{4}= 67.53 $, $ \tau_{4} = 1987.01 $  & \\     
      \hline     
             &   &  &  & $E_{\infty} = 921.77 $, $E_{1} = 118.23 $, $ \tau_{1} = 7.15 $  & \\
             &  25   & 0.02  & 4  & $E_{2} = 127.92 $, $\tau_{2}=61.44 $, $ E_{3} = 120.61 $  & 0.485\\
             &     &      &   & $\tau_{3} = 372.49 $, $ E_{4}= 174.36 $, $ \tau_{4} = 2886.36 $  & \\ 
             \cline{2-6}
RefelektLean &   &  &  & $E_{\infty} = 703.55 $, $E_{1} = 123.26 $, $ \tau_{1} = 8.075 $  & \\
             &  45   &  0.02    & 4  & $E_{2} = 152.68 $, $\tau_{2}=70.38 $, $ E_{3} = 99.89 $  & 0.843\\
             &     &      &   & $\tau_{3} = 340.73 $, $ E_{4}= 157.29 $, $ \tau_{4} = 2096.87 $  & \\  
             \cline{2-6}
             &   &  &  & $E_{\infty} = 466.78 $, $E_{1} = 120.22$, $ \tau_{1} = 6.97 $  & \\
             & 65    &  0.02    & 4  & $E_{2} = 121.301 $, $\tau_{2}=54.067 $, $ E_{3} = 72.56 $  & 1.53\\
             &     &      &   & $\tau_{3} = 273.29 $, $ E_{4}= 85.613 $, $ \tau_{4} = 2298.70$  & \\ 
             \cline{2-6}
             &   &  &  & $E_{\infty} = 267.90 $, $E_{1} = 63.677  $, $ \tau_{1} = 4.27 $  & \\
             & 85    &  0.02    & 4  & $E_{2} = 77.15 $, $\tau_{2}= 20.68 $, $ E_{3} = 60.66 $  & 1.085\\
             &     &      &   & $\tau_{3} = 96.44$, $ E_{4}=  25.63  $, $ \tau_{4} = 1423.905 $  & \\   
      \hline     
             &  & & & $E_{\infty} = 1355.09  $, $E_{1} = 95.945 $, $ \tau_{1} = 9.38 $  & \\
             & 25    &  0.0417    & 5  & $E_{2} = 92.683 $, $\tau_{2}= 85.95 $, $ E_{3} = 78.51 $  & 0.3318\\
             &     &      &   & $\tau_{3} = 534.650 $, $ E_{4}= 80.385 $, $ \tau_{4} = 2891.94 $  & \\   
             &     &      &   & $ E_{5}= 133.992 $, $ \tau_{5} = 25430.43 $  & \\      
             \cline{2-6}
     Lexan   &   &  &  & $E_{\infty} = 1160.15 $, $E_{1} = 102.074 $, $ \tau_{1} = 7.486 $  & \\
             & 45    &  0.0417    & 4  & $E_{2} = 148.61 $, $\tau_{2}=99.854 $, $ E_{3} = 117.735 $  & 3.40\\
             &     &      &   & $\tau_{3} = 901.367 $, $ E_{4}= 126.03 $, $ \tau_{4} = 6086.96 $  & \\  
             \cline{2-6}
             &   &   &  & $E_{\infty} = 996.04 $, $E_{1} = 153.49$, $ \tau_{1} = 11.84 $  & \\
             &   65  &  0.0417    &  4 & $E_{2} = 122.07 $, $\tau_{2}=96.92 $, $ E_{3} = 142.25 $  & 0.9792\\
             &     &      &   & $\tau_{3} = 860.72 $, $ E_{4}= 147.93 $, $ \tau_{4} = 5758.89 $  & \\     
             \cline{2-6}
             &   & & & $E_{\infty} = 603.52 $, $E_{1} = 170.69  $, $ \tau_{1} = 10.88 $  & \\
             &  85   &   0.0417    & 5  & $E_{2} = 165.59 $, $\tau_{2}= 93.78 $, $ E_{3} = 209.43 $  & 1.366\\
             &     &       &   & $\tau_{3} =  686.34 $, $ E_{4}=  169.95  $, $ \tau_{4} = 1976.20 $  & \\   
             &     &      &   & $ E_{5}= 125.12 $, $ \tau_{5} =  24459.21 $  & \\               

            %& 105  & 3 &   &  \\
      \hline      
       \end{tabular}
    \caption{Material parameters identified based on prony series (PS) model.}
    \label{PSmodel}
\end{table}

\begin{table}[h!]
    \centering
    \begin{tabular}{c c c c c c }
    \hline
     Material & Temperature & Strain&$\alpha$ & A   & RSME  \\
       & (\textdegree C )& $\varepsilon $  && (MPa s$^{\alpha}$)  &  Error $(\%)$\\
     \hline   
          & 25  &0.02 & 0.06858  & 429.133  & 0.5364\\
     OSBS & 45  &0.02 & 0.06390 &  259.943 & 0.6485 \\
      \hline
            & 25  & 0.02 & 0.03110   &  2013.616     & 0.169  \\
    TPE-HD  & 45  & 0.02 & 0.054732 & 1783.832   & 0.155  \\
            & 65  & 0.02 & 0.106298 & 1503.640   &  0.558\\
            & 85  & 0.02 & 0.084548 & 757.535    & 0.828  \\
      \hline     
             & 25  & 0.02 & 0.059327 & 1643.10  &   0.3313 \\
ReflektLean & 45  & 0.02 & 0.074289 & 1412.195 &  0.3329    \\
             & 65  & 0.02 & 0.067308 &  875.67  &  0.407 \\
             & 85  & 0.02 & 0.044636 & 397.10   & 1.067 \\
      \hline     
            & 25  & 0.0417 &  0.030154 & 1946.379  & 0.1412 \\
            & 45  & 0.0417 &  0.042348  &  1791.409&  0.2087\\
    Lexan   & 65  & 0.0417 &  0.054757 & 1749.269   & 0.3227 \\
            & 85  & 0.0417 &  0.099548 &  1833.933 & 1.3014 \\
            %& 105  & 3 &   &  \\
      \hline   
    \end{tabular}
    \caption{Material parameters identified based on fractional calculus model.}
    \label{FMmodel}
\end{table}

\subsection{Parameter estimation for strain-dependent relaxation response at constant temperature}
\begin{figure}[h!]
	\centering
	\includegraphics[width=0.58\linewidth]{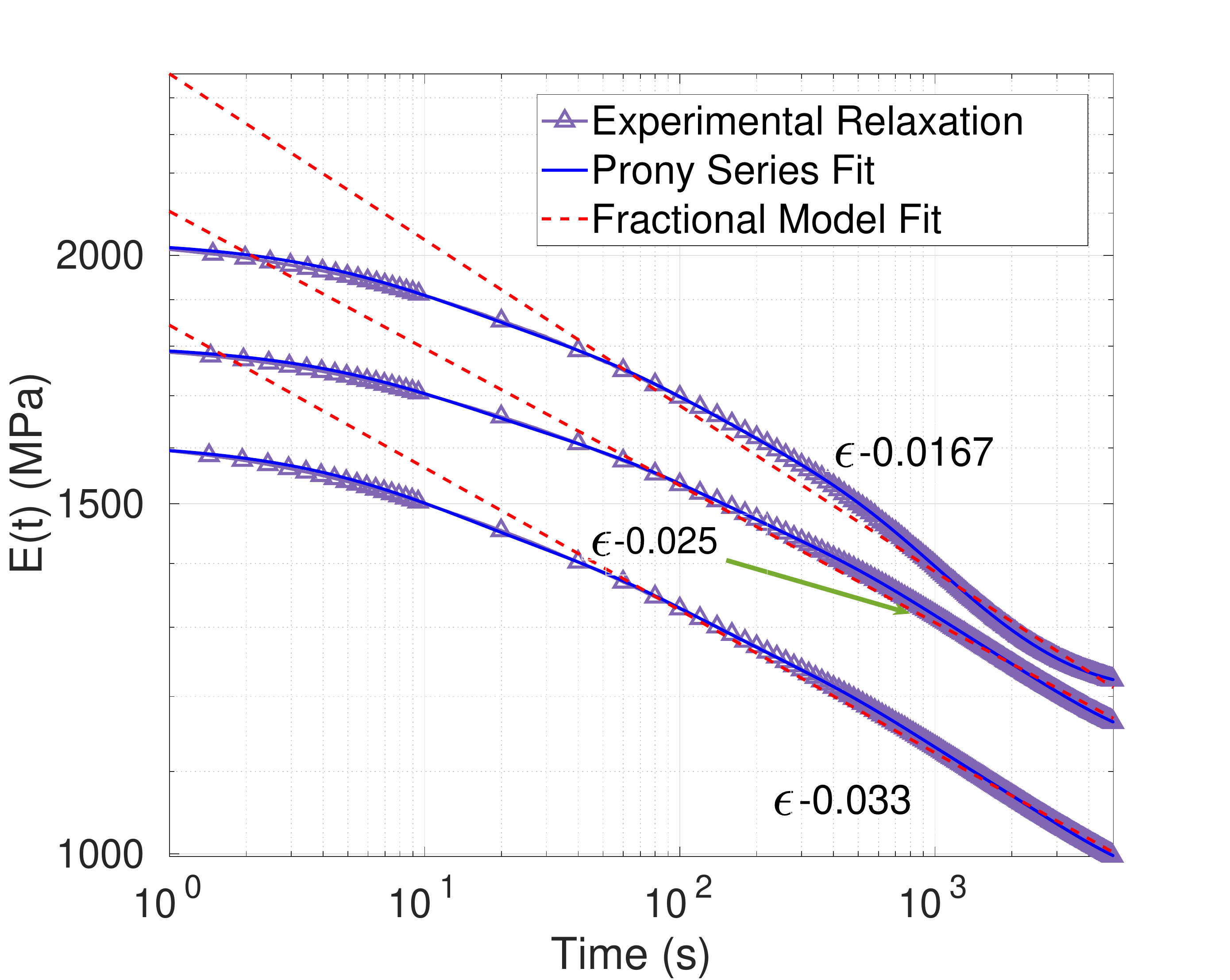}
	\caption{Plot for relaxation response $E(t)$ of Lexan at 85 \textdegree C for Maxwell and Fractional Model for different strains.}
	\label{fig:Lexan_const85}
\end{figure}

The relaxation tests were performed on Lexan back sheet specimens keeping the constant temperature at different strain levels to understand the effect of strains on the relaxation curve.
Fig. \ref{fig:Lexan_const85} shows the relaxation response modeled using PS and FC models against constant strains $\varepsilon_{c} = 0.0167, 0.0250$, and 0.0333, for Lexan at 85 \textdegree C. 
The model parameters for both material models are obtained for all backsheet materials based on the relaxation response obtained from experiments as shown in Tab. \ref{tab:Leaxan_const85}. 
It is noted that at 85 \textdegree C, specimens loaded at small strains relax early compared to those loaded at higher strains. 
No significant change in $\alpha$ was observed, so one may take a constant relaxation pattern for all strains at a constant temperature for Lexan.
In most numerical simulations, taking constant $\alpha$ describing relaxation behavior for a given temperature is helpful. 
Similarly, for the PS series model, which is modeled with 4 arms, relaxation time $\tau_{i}$ and elastic stiffness $E_{i}$ of respective arms seem to be in the same range, which can be taken as constant.
\begin{table}[h!]
    \centering
    \begin{tabular}{cccccccc}
           \hline
                &       &           & & $E_{1}$, & $E_{2}$, & $E_{3}$, & $E_{4}$,  \\
       Strain   &  A       &  $\alpha$ & $E_{0}$ & $\tau_{1}$ & $\tau_{2}$ & $\tau_{3}$ & $\tau_{4}$  \\
                  &      &  &   & \multicolumn{4}{c}{ (MPa, s)}  \\
       \hline
       0.0167 & 2605.01 & 0.0834 &1211.33 & 146.03  & 190.78& 226.27 & 261.85\\
              &   & &        & 9.08 & 73.87 &  593.06 & 1629.53 \\   
              \hline
       0.0250 & 2198.51 & 0.0689 &1131.55 & 122.84 &151.76 & 146.55 & 251.65\\
              &  &  &       &  9.84  & 79.35 & 480.78 & 2477.23\\   
              \hline
       0.0330 & 1930.74 & 0.0716&966.17   & 129.15 & 153.35 & 139.36 & 223.35 \\
              &   &  &        & 9.09   & 70.75  & 438.32 & 2562.24  \\
        \hline
    \end{tabular}
    \caption{Parameter estimation for relaxation response of Lexan at 85 \textdegree C for Maxwell and Fractional Model for different strains}
    \label{tab:Leaxan_const85}
\end{table}

\section{Numerical Analysis}
\label{numericalana}
This section implements models with parameters identified as in the previous section into FEAP as user elements.
They are applied here to some relevant case studies to prove their effectiveness in capturing experimental trends.
Moreover, the numerical models are exploited to simulate realistic conditions for a PV module. Namely, we consider bending tests to assess how the viscoelastic properties of the backsheet affect the axial stress $\sigma_{z}$ along the PV module cross-section. 

\subsection{Numerical validation of relaxation tests}
\label{RelaxationRes}
\begin{figure}[h!]
    \centering
    \includegraphics[width=1.\linewidth]{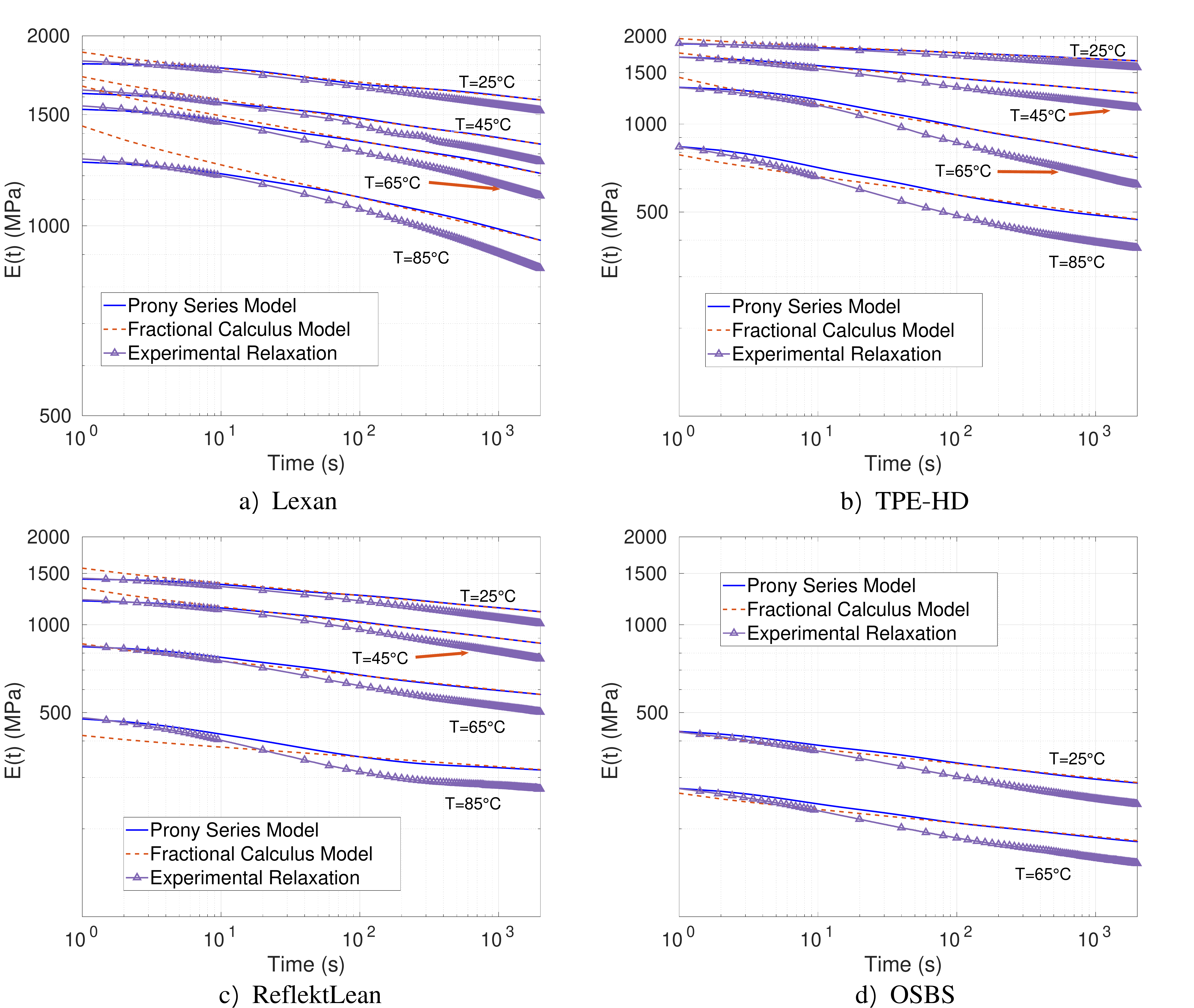}
    \caption{Numerical results obtained from the simulations for elastic relaxation modulus $E(t)$ versus time.}
    \label{fig:NumLe/TPE}
\end{figure}

Numerical simulation of relaxation tests was performed on specimens of dimensions 50 mm $\times$ 20 mm $\times$ thickness of the respective backsheets. 
The bottom side is fixed, while and the top side is subjected to constant strain so that $\epsilon_{0} = d/l_{0}$. 
The considered strain data for each material are collected in Tab. \ref{Erelaxd}. 
The inertia of the material is neglected. Material parameters have already been estimated from experimental relaxation tests on different backsheets in the previous section \ref{Paraidentify}.
Other input parameters besides the relaxation modulus are the Poisson's ratio $\nu = 0.30$ and the time-independent bulk modulus $K = E/3(1-2 \nu)$. 
The PS and FC models show good convergence for a fixed time increment $\Delta t = 1$ s.
Fig. \ref{fig:NumLe/TPE} shows the numerically predicted elastic relaxation response of backsheet materials as compared to  experimental results. 
The elastic relaxation modulus can be evaluated as $E(t) = \sigma(t)/ \epsilon_{0}$, where $\sigma(t)$ is the nominal stress at time $t$ predicted by numerical simulations.
The numerical results for different backsheet materials show good agreement with the experimental tests. The slight deviations from experimental results can be ascribed to a time-independent bulk modulus assumption. 

\subsection{Effect of the initial loading rate on creep and relaxation experiments}
\begin{figure}[h!]
    \centering
    \includegraphics[width=1\linewidth]{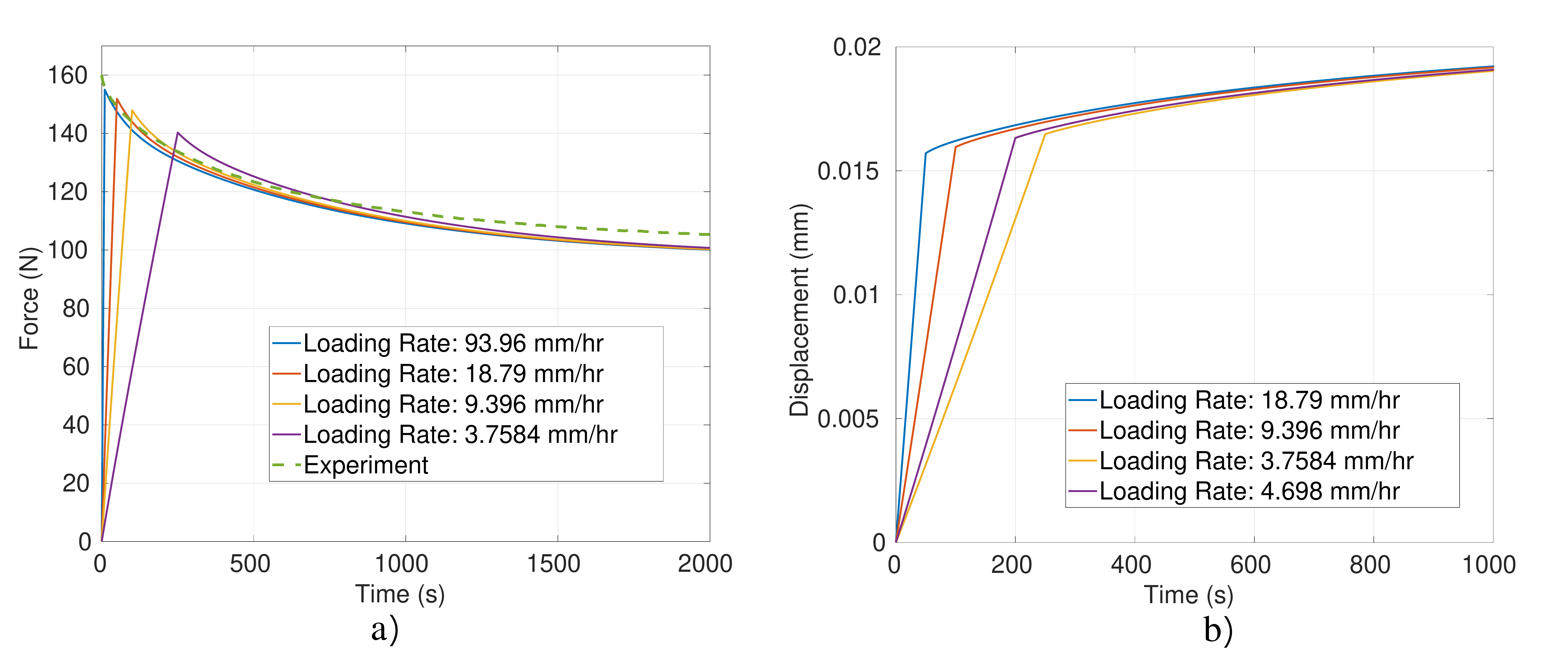}
    \caption{Numerical results from simulations for the effect of strain rates on a) Relaxation and b) Creep response of Lexan backsheet at 85 \textdegree C considering 5 Maxwell arms.}
    %\textcolor{red}{specify that the results are numerically obtained with Maxwell with 5 arms. Not LR, write loading rate otherwise is difficult to follow}
    \label{StrainRateff}
\end{figure}

\begin{figure}[h!]
    \centering
    \includegraphics[width=0.75\linewidth]{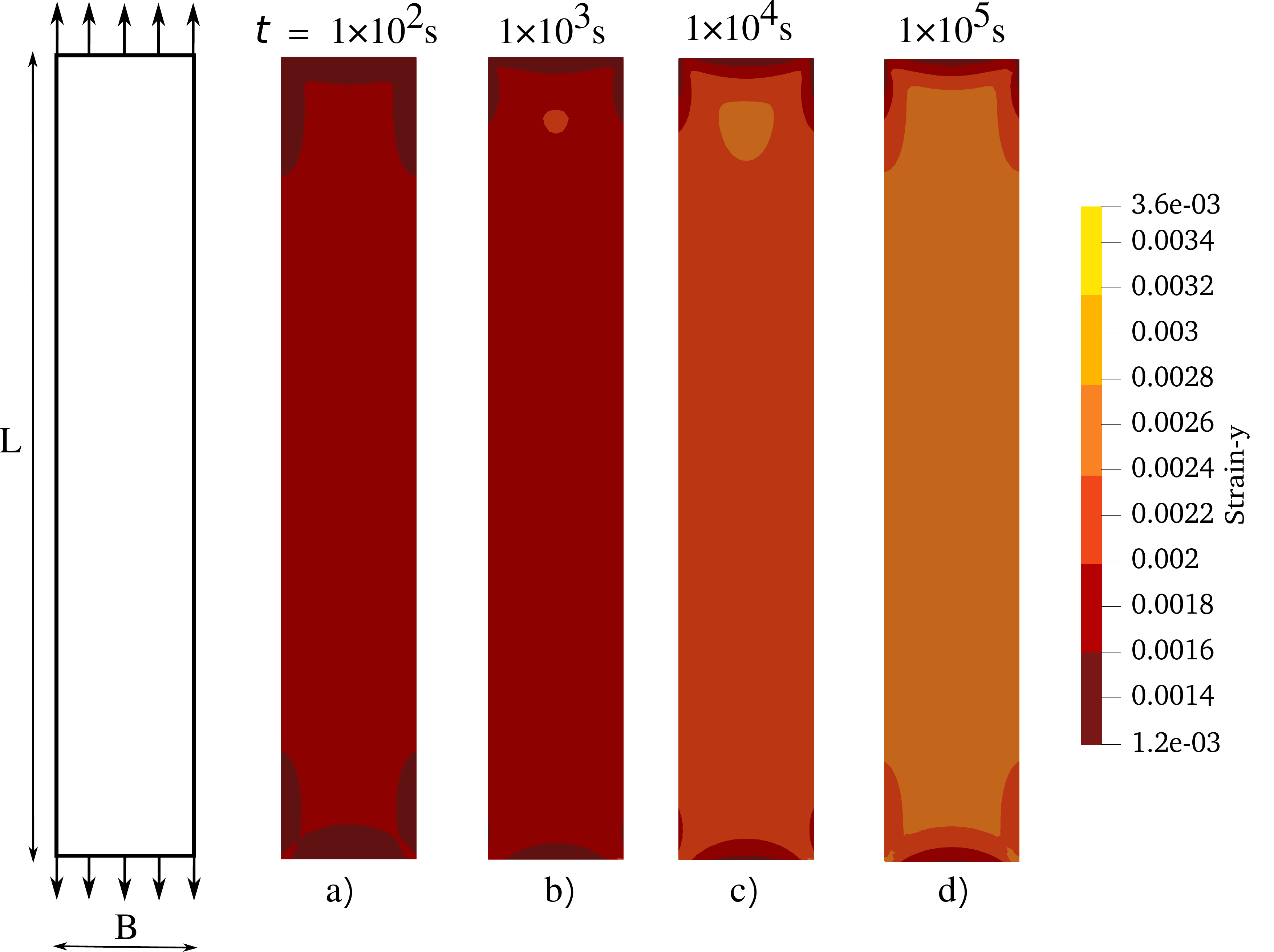}
    \caption{Simulation of Lexan backsheet creep experiment at 25 \textdegree C: contour plot of the tensile strain at different points and its evolution in time.}
    \label{fig:Lexan_Creep}
\end{figure}

To understand the effect of the initial loading rate on elastic relaxation and creep of materials, a numerical study is performed with a different initial loading rate for the Lexan backsheet. 
A specimen of dimensions 60 mm $\times$ 40 mm $\times$ 0.296 mm is clamped on the bottom side and loaded on top with a constant strain $\varepsilon = 0.0065 $, over the period of the relaxation test, 
while a constant force $F = 30$ N is maintained over time for the creep tests. 
The numerical results with the 5 Prony Series arm model are shown in Fig. \ref{StrainRateff}. 
It can be observed that the initial loading rate significantly affects the initial stress $\sigma_{0}$ or strain $\varepsilon_{0}$. 
But, over time, both relaxation and creep responses with varying initial loading rates converged together as we kept all input material parameters constant.
From those plots, it is evident that the initial rate of loading does not affect the long-term response. 
Fig. \ref{fig:Lexan_Creep} shows a specimen's numerically predicted creep behavior at 25 \textdegree C with dimensions $L= 120 $ mm, $B = 20 $ mm, and 0.296 mm thick Lexan backsheet subjected to the constant 
$F = 40$ N. The contour plot of the axial strain $\epsilon_{l}$ over a time period from 100 to 10,000 s is shown. The predicted spatial variation of strain is the same in both viscoelastic models.

\subsection{Response under creep cycles}
\begin{figure}[h!]
    \centering
    \includegraphics[width=0.95\linewidth]{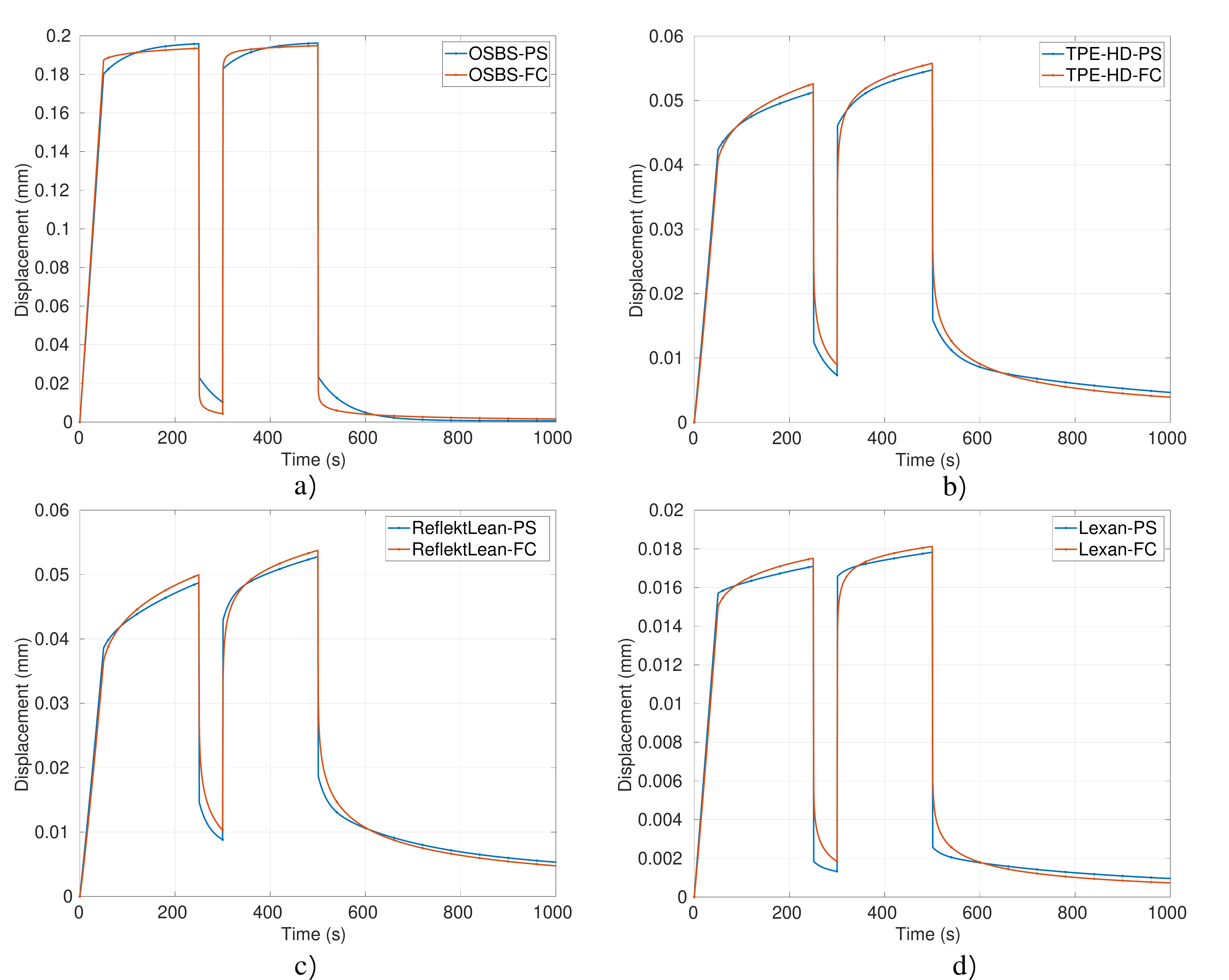}
    \caption{Response of backsheet materials a) OSBS, b) TPE-HD, c) ReflektLean, and d) Lexan under creep cyclic loading at 25 \textdegree C.}
    \label{CreepRes}
\end{figure}

A numerical investigation is performed to understand the response of backsheet materials, with parameters identified in Tab. \ref{PSmodel}, when subjected to creep cycles as in Fig.\ref{CreepRes}. 
A specimen of dimensions 60 mm $\times$ 40 mm $\times$ thickness is clamped on the bottom side and loaded on top. All materials were subjected to the constant traction $F = 30$ N over the first 200 s, relaxed for the next 50 s, and again loaded with constant $F = 30$ N for another 200 s period, followed by unloading for the rest of the time.
The response of both material models, and for all the backsheet materials, is shown in Fig.\ref{CreepRes}. 
Both PS and FC material models show good predictions. It also shows that FC models with only two unknown parameters are able to predictable responses as PS models with a large number of unknowns.
As OSBS has a lower elastic modulus than the other materials, it deforms more than the others, and it also gets faster relaxed. 
An important observation that can be made from all the material responses is that, after the first cycle of loading, the material deforms more at the next cycle for the same load, as compared to the first one. 
This is primarily due to the polymeric response of backsheet materials, as chains of constituent polymers get unstrained and stretched over the first cycle, followed by relaxation. 
This response is very important in the case of failures of PV modules that are repeatedly subjected to cyclic loads. 

\subsection{Three-point bending of a PV mini-module}
\label{3PTbending}
\begin{figure}
    \centering
    \includegraphics[width=1\linewidth]{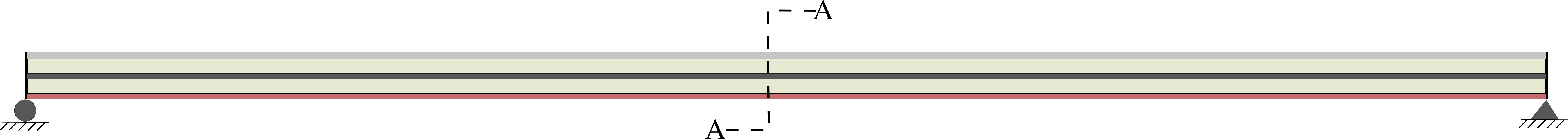}
    \caption{Schematic showing three-point bending of a PV mini-module.}
    \label{Bending_Exp}
\end{figure}

\begin{figure}[h!]
    \centering
    \includegraphics[width=0.8\linewidth]{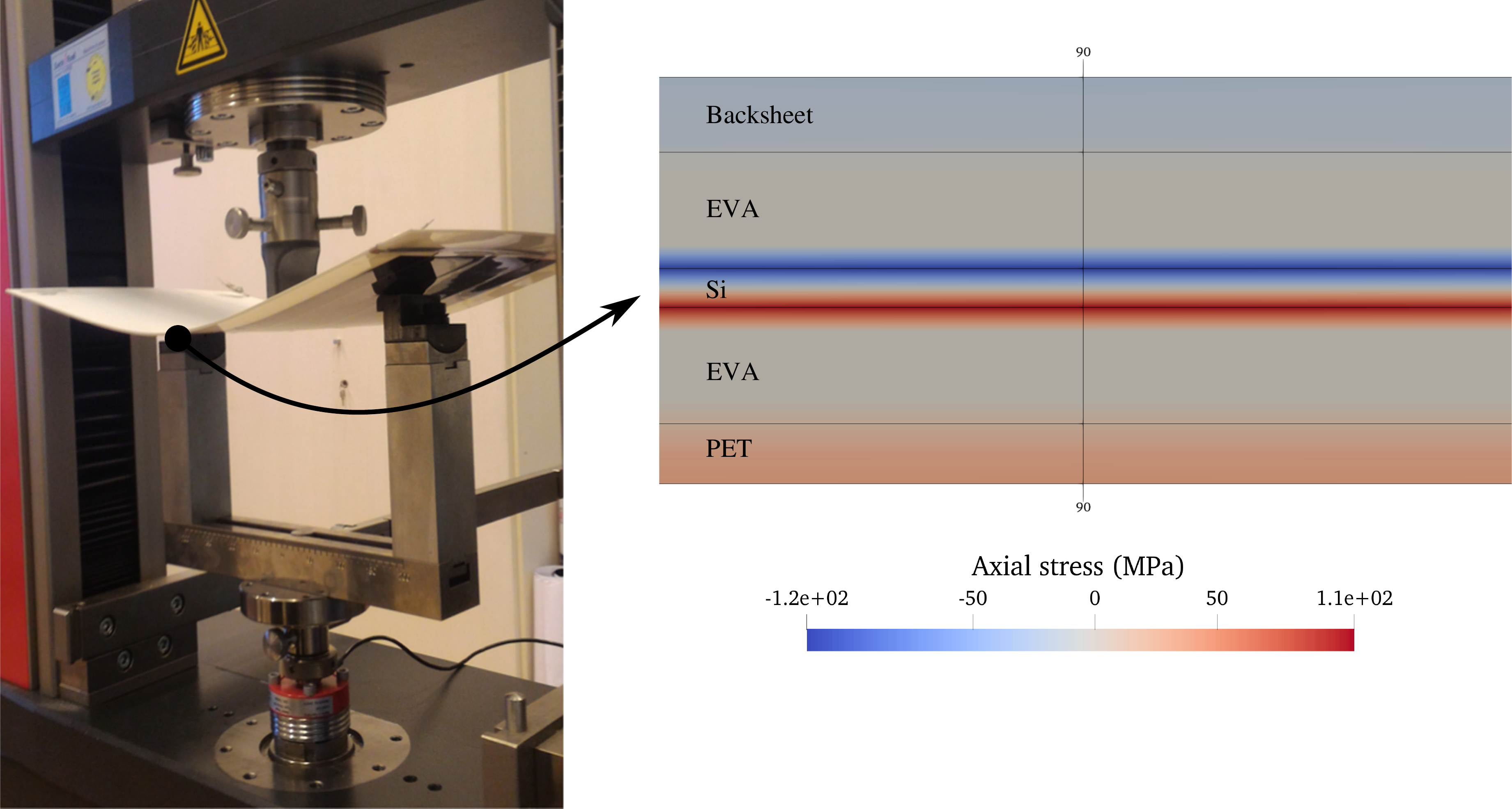}
    \caption{a)The experimental setup for the bending test for PV laminate \citep{borri2018fatigue}, and b) Axial stress $\sigma_{z}$(MPa) across the mid-cross-section of the PV mini-module under constant displacement $u_{y} = -20 $mm with Lexan backsheet at 85 \textdegree C.}
    \label{FEA3PT}
\end{figure}

\begin{figure}[h!]
    \centering
    \includegraphics[width=1\linewidth]{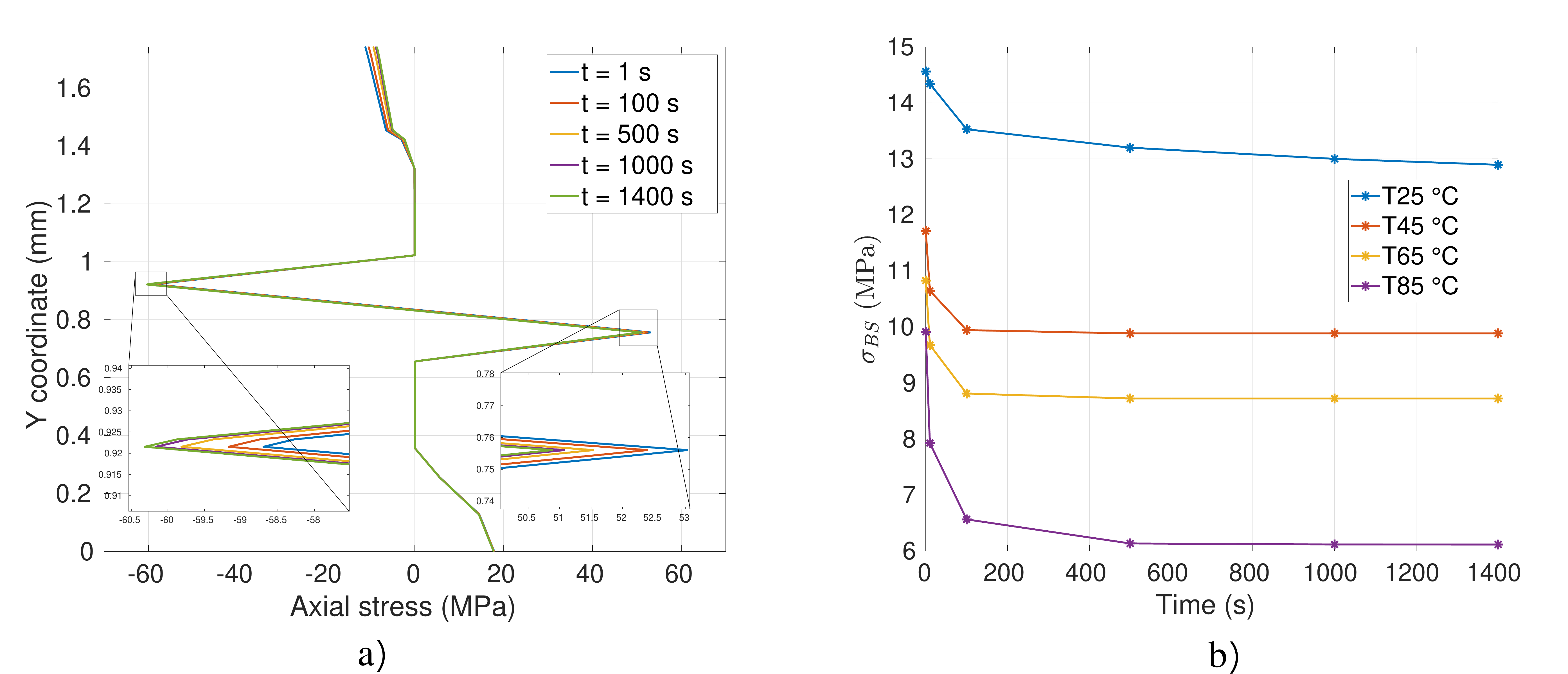}
    \caption{a) Axial stress $\sigma_{z}$(MPa) across the mid-cross-section of the PV mini-module under constant displacement $u_{y} = -20 $mm with Lexan backsheet at 85 \textdegree C, and
    b) Variation of $\sigma_{BS}$ (MPa) with respect to time over a different ranges of temperature when subjected to bending}
    \label{Bend12}
\end{figure}

In this section, using the identified model parameters given in Tab. \ref{PSmodel}, the response of a PV mini-module under bending is being tested for a long duration of time to understand the effect of viscoelastic properties of the backsheet in terms of axial stresses $\sigma_{z}$ developed within the PV module. 
The PV mini-module under study is composed of the following layers (from the intrados to the extrados): a backsheet 0.32 mm thick, an EVA layer with $E_{\text{EVA}} = 10$ MPa  0.5 mm thick, a polycrystalline Si solar cell with $E_{\text{Si}} = 130$ GPa  0.166 mm thick, another EVA layer 0.5 mm thick, and finally a PET with $E_{\text{PET}} = 2800$ MPa cover 0.265 mm thick. The span of the mini-module is $L = 180 $ mm. 
The three-point bending experiment and boundary conditions are schematically shown in Fig.\ref{Bending_Exp}.
Following the experimental tests performed by \citep{borri2018fatigue}, the PV mini-module was subjected to a maximum deformation of 20 mm at its mid-span position, which led to a reaction force of 64 N.
The experimental setup is shown in Fig. \ref{FEA3PT}. In the FE analysis, Poisson's ratio $\nu$ is taken as 0.3 for all layers, assuming that all layers are perfectly bonded. 
The time-dependent response of backsheet materials is introduced based on the elastic moduli values at 85\textdegree C, as given in Tab. \ref{Ecompare}.
The PV mini-module is subjected to a constant mid-span defection of $u_{y} = -20 $ mm over the whole time period. 
As correspondence principle \citep{alotta2018finite} proved to be valid for linear viscoelasticity, the shape of the deformed configuration of the backsheet layer is the same as that with elastic one only difference being stress across the layer, which scales with time and directly affects stress configuration across whole PV module with time.
The axial stress $\sigma_{z}$ distribution for the PV module considering Lexan as backsheet material over time is shown in Fig. \ref{FEA3PT}.
 The variation of stress $\sigma_{BS}$, over time across a range of temperatures is shown in Fig.\ref{Bend12}. It shows that as $\sigma_{BS}$ decreases as temperature increases. 
It can be observed that with the reduction of $E(t)$ over time, higher stresses tend to build across the Si layer.
The timely degradation of elasticity may lead to mechanical failure of Si cells, which are too thin and brittle than other layers.

\section{Discussion and conclusion}
\label{conclusion}
\begin{figure}[h!]
    \centering
    \includegraphics[width=1\linewidth]{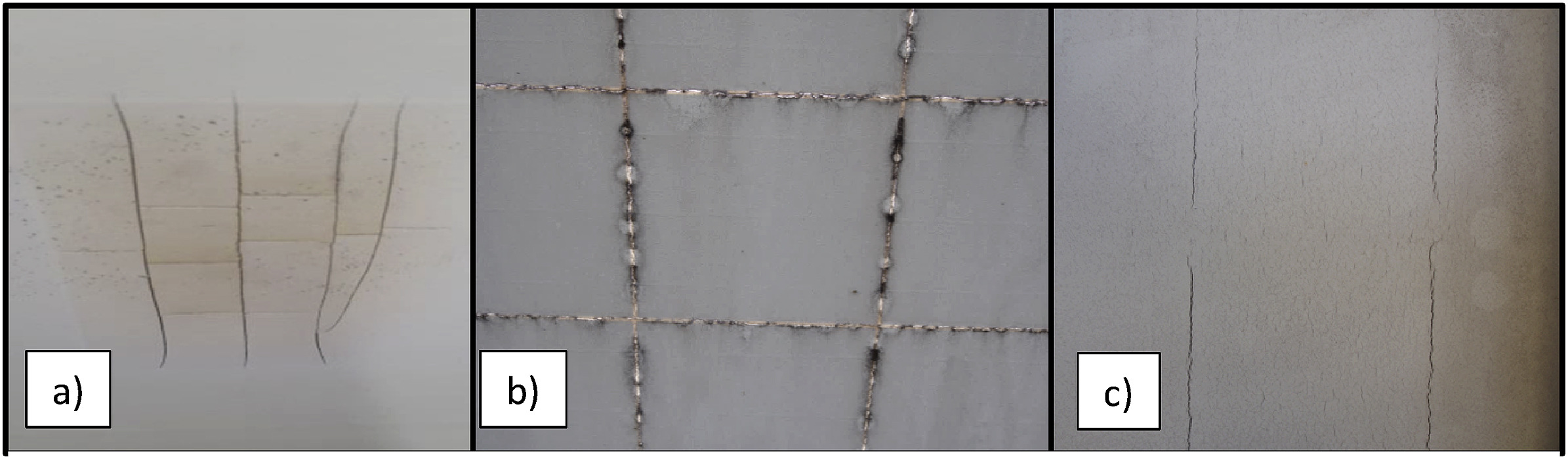}
    \caption{Common failures in PV backsheet due to thermo-mechanical and material degradation: a) cracked backsheet beneath a hot Si-cell, b) squared cracks beneath cell interspaces, and c) longitudinal cracks located beneath bus-bars, from \citep{eder2019error}.}
    \label{fig:solarfail}
\end{figure}

In this study, the viscoelastic response of commercially available backsheet materials is experimentally characterized for temperature-dependent relaxation and uniaxial tensile tests. 
An extensive viscoelastic experimental study on backsheet materials is carried out, considering the temperature-dependent properties for characterizing the mechanical properties. 
Experiments show that the Lexan backsheet has the highest strength compared to other back sheet materials under study.
Based on an experimental campaign, we have proposed small-strain viscoelastic models based on the Prony-series (PS) and Fractional Calculus (FC) models for constitutive equations.
We have implicitly implemented both models using user element (UEL) subroutines in the FEAP \citep{taylor2014feap}. 
We have shown that the FC model can be implemented in finite elements by using the discretized version of the fractional derivative provided by Grunwald-Letnikov (GL) and the PS model through a semi-analytical approach. The current element formulation can be easily implemented with any FE code. 
A robust optimization procedure for identifying viscoelastic material parameters for both models.
It was observed that the FC model needs only two parameters, $A$ and $\alpha$, to characterize the relaxation response,  while the PS model needs the identification of 7 to 11 unknown parameters. 
It is also noted that FC models cannot predict the relaxation behavior for very short periods under $10^{1}$ s. Still, they give very good predictions for the long-term response, which is the regime of interest for applications.
FC models require the complete time history of strains compared to the PS model, in which only the strain at a current and a previous time step is required. 
In this regard, the FC model needs more memory to store the history variables as compared to the PS model. 
\\
The parameters identified can be directly used as a material input for simulating the response of the respective backsheet materials. 
Following identifications of relevant material parameters, we have validated the model with the experimental data that shows good predictability. 
From the numerical experiments, we also showed that when the material is loaded under creep cyclic loads, the material undergoes more deformation as long as the number of cycles increases. 
Such an extensive experimental study and constitutive modeling will help design and simulate a more comprehensive modeling of PV modules.
To understand the effect of the viscoelastic response of the backsheet on the stress field inside a PV module, a three-point bending experiment was simulated with parameters identified for both viscoelastic material models. 
Numerical simulations show that the elastic modulus reduction over time of the backsheet material leads to an increase in the axial stress $\sigma_{z}$ to be withstood by the Silicon layer, which may enhance the brittle failure of Silicon solar cells.
Future investigations will concern the fracturing of the PV-back sheet, as shown in Fig. \ref{fig:solarfail}.
In most cases, these failures occur due to material degradation over time. 
To address fracture in the backsheet, the present framework can  be coupled with a phase field approach to fracture, which could be effectively applied to simulate and capture such complex material-degradation scenarios.

\section{Acknowledgements}
\noindent
The authors acknowledge funding from the Italian Ministry of University to the research project of national interest PRIN 2017 XFAST-SIMS: Extra fast and accurate simulation of complex structural systems (MUR code 20173C478N).

%Bibliography
\bibliography{references}

\end{document}